
%
%
%
%
%
%
%
\input harvmac
\def\figflag{I}
%
\global\newcount\figno \global\figno=1
\newwrite\ffile
\def\tfig#1{Fig.~\the\figno\xdef#1{Fig.~\the\figno}\global\advance\figno by1}
\def\figI{I}
%
\newdimen\tempszb \newdimen\tempszc \newdimen\tempszd
\newdimen\tempsze
\ifx\figflag\figI
\input epsf
\def\epsfsize#1#2{\expandafter\epsfxsize{
 \tempszb=#1 \tempszd=#2 \tempsze=\epsfxsize
     \tempszc=\tempszb \divide\tempszc\tempszd
     \tempsze=\epsfysize \multiply\tempsze\tempszc
     \multiply\tempszc\tempszd \advance\tempszb-\tempszc
     \tempszc=\epsfysize
     \loop \advance\tempszb\tempszb \divide\tempszc 2
     \ifnum\tempszc>0
        \ifnum\tempszb<\tempszd\else
           \advance\tempszb-\tempszd \advance\tempsze\tempszc \fi
     \repeat
\ifnum\tempsze>\hsize\global\epsfxsize=\hsize\global\epsfysize=0pt\else\fi}}
\epsfverbosetrue
\fi
%
%
\def\ifigure#1#2#3#4{
\midinsert
\vbox to #4truein{\ifx\figflag\figI
\vfil\centerline{\epsfysize=#4truein\epsfbox{#3.eps}}\fi}
\narrower\narrower\noindent{\bf #1:} #2
\endinsert
}

\def\bz{{\bar z}}
\def\del{\partial}
\def\bdel{{\bar\partial}}
\Title{\vbox{\hbox{HUTP-93/A002} \hbox{TIFR/TH/93-01}}}
{\vbox{\centerline{Two Dimensional Black-hole as a Topological}
\medskip
\centerline{Coset Model of c=1 String Theory} }}

\bigskip
\bigskip

\centerline{Sunil Mukhi}
\medskip\centerline{Tata Institute of Fundamental Research}
\centerline{Homi Bhabha Rd}
\centerline{Bombay 400 005, India}
\bigskip\bigskip

\centerline{Cumrun Vafa}
\medskip\centerline{Lyman Laboratory of Physics}
\centerline{Harvard University}
\centerline{Cambridge, MA 02138, USA}

\vskip .3in
We show that a special superconformal coset (with $\hat c =3$) is
equivalent to $c=1$ matter coupled to two dimensional gravity.  This
identification allows a direct computation of the correlation functions of
the $c=1$ non-critical string to all genus, and at nonzero cosmological
constant, directly from the continuum approach. The results agree with
those of the matrix model. Moreover we connect our coset with a twisted
version of a Euclidean two dimensional black hole, in which the ghost and
matter systems are mixed.

\Date{1/93}

\newsec{Introduction}
Two dimensional quantum field theories coupled to gravity have been
studied in great detail in the last few years (see the review article
\ref\mgin{P. Ginsparg and G. Moore,
Proceedings of TASI 92, ed. J. Harvey and
J. Polchinski.}). In particular the conformal theories
for which the central charge is less than or equal to 1 coupled to gravity
are believed to make sense on arbitrary genus. The case $c=1$ corresponds
to the point where the tachyon is just massless, and beyond that we have a
tachyonic theory which is not expected to make sense at all genus (i.e.,
there must be a kind of phase transition).

It is natural therefore to consider the interesting case of $c=1$ string
theory. This has been studied in the matrix model formalism beginning
with \ref\mace{E. Brezin, V. Kazakov and Al.B. Zamolodchikov, Nucl. Phys.
B338 (1990) 673\semi D. Gross and N. Miljkovic, Phys. Lett. 238B
(1990) 217\semi P. Ginsparg and J. Zinn-Justin, Phys. Lett. 240B
(1990) 333\semi G. Parisi, Phys. Lett. 238B (1990) 209.}\ and
considerably elaborated in subsequent works (see for example
\ref\klebrev{I. Klebanov, {\it ``String Theory in Two Dimensions''},
Preprint PUPT-1271, July 1991\semi
D. Kutasov, {\it Some properties of (non)critical strings},
Lectures given at ICTP Spring School on String Theory and Quantum
Gravity, Trieste, Italy 1991.} and references therein). There is also a
collective field theory description of the target space \ref\da{S. Das and
A. Jevicki, Mod. Phys. Lett. A5 (1990) 1639.}\ as well as a
Kontsevich-like description (using a generalized Penner model)
\ref\dijmoor{R. Dijkgraaf, G. Moore and R.  Plesser,{\it The
partition function of 2D string theory}, IASSNS-HEP-92/48, YCTP-P22-92.}.

Some results have also been obtained from the continuum viewpoint, using
the well-known DDK formalism \ref\ddk{F. David, Mod. Phys. Lett. A3 (1988)
1651\semi J. Distler and H. Kawai, Nucl. Phys. B321 (1989) 509.}.
This includes the computation of correlators on the sphere
\ref\dikut{J. Polchinski, Nucl. Phys. B 362 (1991) 125\semi
P. DiFrancesco and D. Kutasov, Phys. Lett. 261B (1991) 385}\ and the partition
function on the torus
\ref\bekl{M. Bershadsky and I. Klebanov, Phys. Rev. Lett. 65 (1990) 3088.},
and the classification of physical states \ref\lianetc{ B. Lian and G.
Zuckerman, Phys. Lett. 266B (1991) 21\semi S. Mukherji, S. Mukhi
and A. Sen, Phys. Lett. 266B (1991) 337\semi P. Bouwknegt, J.
McCarthy and K. Pilch, Comm. Math. Phys. 145 (1992) 541.}.

Two recent discoveries have uncovered new and interesting aspects of $c=1$
string theory. The first is the discovery of a 2d area-preserving symmetry
algebra and a natural ring of fields \ref\wittring{E. Witten, Nucl. Phys.
B373 (1992) 187\semi E. Witten and B. Zwiebach, Nucl. Phys. B377
(1992) 55.}\ref\pol{I. Klebanov and A. Polyakov, Mod. Phys. Lett. A6
(1991) 3273.}\ which gives rise to powerful Ward-identities \ref\kleb{I.
Klebanov, Mod. Phys. Lett. A7 (1992) 723.}. The second is the observation that
the $SL(2,R)/U(1)$
coset, which can be viewed as a two dimensional black-hole, is very
similar to the $c=1$ model \ref\bwit{E. Witten, Phys. Rev. D44 (1991)
314.}\ and this has given rise to the hope that exact computations using
the matrix model may shed light on features of two-dimensional black
holes (for some recent work in this direction, see \ref\dasetc{S. Das,
Preprint TIFR/TH/92-62, October 1992\semi A. Dhar, G. Mandal and
S. Wadia, Preprint TIFR/TH/92-63, October 1992.})

Although some matrix-model results have been re-derived in the continuum
DDK formalism, the latter has major limitations, in that it is practicable
only in low genus, and one requires certain prescriptions to handle the
case of nonzero cosmological constant. It would be highly desirable to
have a more powerful continuum formulation in which higher-genus and
non-zero cosmological constant are as natural as in matrix models. It has
frequently been suggested that a topological field theory could provide
the relevant exactly solvable continuum formulation, but so far it has
remained an open problem to find the right topological theory.

Two additional results suggest a possible solution to this open problem.
The first \ref\DV{J. Distler and C. Vafa, Mod. Phys. Lett. A6 (1991)
259.}\ is that the $c=1$ string at the self-dual radius computes the Euler
characteristic of the moduli space of Riemann surfaces, and can be
represented by the double scaling limit of the Penner model
\ref\pen{R.C. Penner, J. Diff. Geom. 27 (1988) 35.}. The second is the
observation \ref\wittks{E. Witten, Nucl. Phys. B371 (1992) 191.}\ that the
partition function of a twisted $N=2$ superconformal $SU(2)/U(1)$ coset
model at $k=-3$ (with $\hat c=3$) is also given by the Euler
characteristic of the moduli space of Riemann surfaces. This naturally
raises the question of whether there is a direct relation between the two.
Our primary aim in this paper is to show that they are indeed the same
theory. It follows that the coset model provides the desired topological
field theory description of the $c=1$ string.

The motivation to reconsider this correspondence came from the recent
discovery that there is quite generally an $N=2$ superconformal symmetry
in the bosonic string theory \ref\gatosem{B. Gato-Rivera and A.
Semikhatov, Phys. Lett. 288B (1992) 38.}. This was further elucidated and
generalized \ref\bers{M. Bershadsky, W. Lerche, D. Nemeschansky and N.
Warner, {\it A BRST operator for non-critical W-Strings},
HUTP-A034/92.}\ to essentially all other
known string theories. The simplest description of this $N=2$ symmetry in
the bosonic string comes from the observation that the algebra obeyed by
the stress-energy tensor, the BRST current, the $b$ ghost and the ghost
number current, resembles the twisted $N=2$ superconformal algebra.
Indeed, by modifying the BRST and ghost number currents with total
derivative terms involving the Liouville field, one can obtain precisely a
twisted $N=2$ algebra. Our starting point is that in the $c=1$ string, if
instead of the Liouville field one uses the matter field to modify the
BRST operator, the resulting theory has $\hat c=3$.  This suggests a
possible correspondence with the superconformal $SU(2)/U(1)$ model at
level $-3$, which could explain the observations in the previous
paragraph.

The organization of this paper is as follows:  In section 2 we review how
the $N=2$ superconformal symmetry arises in bosonic strings, and how it is
realized in the particular case of $c=1$ matter coupled to gravity at zero
cosmological constant. In section 3 we discuss the special Kazama-Suzuki
(KS) superconformal $SL(2,R)/U(1)$ coset and show that there is a one to
one correspondence between physical states of this model and those of the
$c=1$ theory in the standard DDK approach. In section 4 we prove that this
twisted KS theory coupled to topological gravity is the same as the $c=1$
theory with non-zero cosmological constant, using the KPZ formalism
\ref\kpz{V. Knizhnik, A. Polyakov and A.B. Zamolodchikov, Mod. Phys. Lett.
A3 (1988) 819.}\ which turns out to have the {\it same} Hilbert space as
that of the KS model. In section 5 we show how the computations of
\wittks\ for this coset reproduce various results of $c=1$ matrix models.

In section 6 we discuss the black-hole interpretation of this model. The
main difference between this description and that of \bwit\ is that here
the matter and the ghosts are mixed. We argue why this may be natural and
why the KS coset model should be viewed as the correct description of
bosonic strings propagating in the 2 dimensional black-hole background.
We also make a preliminary study of the implications of this topological
symmetry for the black-hole singularity. In section 7 we present our
conclusions and some further generalizations of these ideas. In this setup
it seems that the natural generalization is in terms of twisted $G/G$
theory \ref\bsadv{M. Bershadsky, V. Sadov and C. Vafa, work in progress.}.

In appendix A, which is due to Edward Frenkel, the cohomology of the
relevant Kazama-Suzuki model is computed, and is also shown to be
equivalent to that of the {\it bosonic} $SL(2,R)/SL(2,R)$ topological theory.

\newsec{c=1 String Theory as N=2 Topological Field Theory}
It has long been thought that the bosonic string in any background
possesses an extended conformal algebra similar to that of the twisted
$N=2$ superconformal algebra, where the generators are the total
stress-energy tensor $T(z)$ of the matter plus ghost system, the BRST
current $G^+(z)$, the antighost $b(z)$, and a $U(1)$ current $J(z)$ which
counts the ghost number. However, with this choice of generators the
algebra does not close, and one has to add two more generators, the ghost
$c(z)$ along with $c\del c(z)$ \ref\ver{R. Dijkgraaf, E. Verlinde and H.
Verlinde, {\it Notes on Topological String Theory and 2D Quantum Gravity},
Preprint PUPT-1217, IASSNS-HEP-90/80 (November 1990).}.

More recently, it was understood \gatosem \bers\ that with certain
modifications, these generators indeed form a twisted $N=2$ algebra which
closes without adding any other generators. The modifications consist of
adding total derivative terms to the BRST current and the ghost number
current. Indeed, if the background has any free scalar field $\eta(z)$
(possibly with a background charge $Q_\eta$) then the following chiral
fields form a twisted $N=2$ superconformal algebra: %
\eqn\topalg{\eqalign{
T(z) &= T^M(z) + T^G(z)\cr
G^+(z) &= c(z) T^M(z) + \half :c(z) T^G(z): +~ x~\del^2 c(z)
+ y~\del(c(z)\del\eta(z))\cr
G^-(z) &= b(z)\cr
J(z) &= ~:cb(z): -~ y~\del\eta(z)\cr}    }
where
\eqn\xy{\eqalign{
x &= \half(3 + Q_\eta y)\cr
y &= \half(-Q_\eta + \sqrt{Q_\eta^2 - 8} )\cr}  }
and the algebra has a topological central charge $\hat c= c/3= 2x$.

In particular, in $c=1$ string theory, one has a Fock space of two free
bosons $X(z, \bz)$ and $\phi(z, \bz)$, the $c=1$ matter field and the
$c=25$ Liouville field respectively, and two free fermions, $b(z)$ and
$c(z)$, which are the ghost fields. We wish to explore the topological
field theory based on the twisted $N=2$ superconformal algebra obtained
from this background following the above procedure. In principle we have
two choices for the scalar field $\eta(z)$ to be used in the algebra.
Choosing the Liouville field, however, leads to a difficulty.  When we
turn on the cosmological constant, the Liouville equation of motion
\eqn\liou{\del\bdel\phi + \mu e^\phi = 0}
means that we cannot consider $\del\phi$ as a holomorphic current any
more. However, in the case of $c=1$ theory we have another option which
works {\it even if the cosmological constant is not zero}. We choose the
$c=1$ matter field $X(z,\bz)$ to be $\eta$ in Eq.\topalg\ above. Since
this field has no background charge, it follows that $x=3/2$, $y=\sqrt2 i$
and $\hat c=3$.

This value of $\hat c$ is ``critical'' in topological field theory
\ref\wittop{E. Witten, Nucl. Phys. B340 (1990) 281.}, and corresponds to
the value for Calabi-Yau 3-fold compactifications. Simply put,
criticality means that the partition function of such theories can be
non-vanishing in every genus, with no need for any insertions.  Note that
the value of $\hat c$ depended only on the background charge $Q_\eta$ of
the scalar field used in the symmetry algebra. Hence it is a completely
general result that ``critical'' topological theories arise whenever the
string theory background has at least one non-anomalous $U(1)$ current.

We now examine how the $N=2$ algebra organizes the states of $c=1$ string
theory. To be specific, we will work with this theory at the self-dual
radius, where the states are classified by affine $SU(2)$, and restrict
our attention in this section to the case where $\mu =0$. In the (chiral)
Fock space of $X,\phi,b,c$ we start with the Fock vacua
\eqn\vac{ |\Phi^\pm_{s,n}\rangle = |\sqrt2 n\rangle_X
\otimes |-i\sqrt2 (1\mp s)\rangle_\phi\otimes c_1|0\rangle_G}
where the subscripts $X,\phi,G$ refer to the matter, Liouville and ghost
Fock spaces, and the first two spaces have vacua labelled by momenta, as
usual. We have restricted the continuous variable $s$ to be non-negative
by explicitly writing its sign, while $n$ runs over half-integers of both
signs.

These Fock vacua are created from the $SL(2,C)$ vacuum by the
operators
\eqn\vacop{ \Phi^\pm_{s,n}(z) = c e^{i\sqrt2 n X(z)}
e^{\sqrt2(1\mp s)\phi(z)} }
For each such vacuum and operator, we have a degenerate partner
\eqn\vactwo{|\Psi^\pm_{s,n}\rangle = c_0 |\Phi^\pm_{s,n}\rangle}
obtained by acting with the $c$-ghost zero mode, and a
corresponding field
\eqn\vacoptwo{\Psi^\pm_{s,n}(z) = \del c(z)
\Phi^\pm_{s,n}(z)}
For the moment we work only with the holomorphic parts of the fields,
although eventually these must be combined with their anti-holomorphic
counterparts with the {\it same} value and sign of $s$, since the
Liouville field is non-compact.

It is easy to check that all these Fock vacua are in fact primaries of the
$N=2$ topological algebra. Formally, the operators $\Phi^\pm$ look like
tachyon vertex operators, but not necessarily on-shell. Their conformal
dimension is $n^2 - s^2$. In the particular case when $n=\pm s$, these
operators are in the cohomology of $G^+_0$, and correspond to the physical
(on-shell) tachyons of the $c=1$ string.

To find out whether all other Fock-space states are secondaries of the
topological algebra, we need to understand the projection map from the
$N=2$ module to the Fock space. This map will fail to be bijective if
there are null vectors in the former which are in the kernel of the
projection. The simplest example shows that this is indeed the case.

Consider the states $|\Phi^+_{1,0}\rangle = c_1|0\rangle$ and
$|\Psi^+_{1,0}\rangle = c_0 c_1 |0\rangle$. Both have conformal dimension
$h = -1$, and are mapped onto each other by the action of $G^+_0$ and
$G^-_0$. The $N=2$ algebra generates in principle 8 secondaries at level 1
above these, by the action of $L_{-1}$, $G^\pm_{-1}$ and $J_{-1}$. It is
easy to check that four of these are annihilated by all the positive modes
of the algebra generators, i.e. they are both primary and secondary, hence
null in the $N=2$ module.  These are $G^-_{-1}(c_1|0\rangle)$,
$G^+_{-1}(c_1|0\rangle) - \left(L_{-1} - J_{-1} \right)(c_0
c_1|0\rangle)$, $L_{-1}(c_1|0\rangle) + G^-_{-1}(c_0 c_1|0\rangle)$ and
$G_{-1}^+ c_0 c_1|0\rangle$.

Now we can make the projection to Fock space. The first of the four null
states maps to the $SL(2,C)$ Fock vacuum $|0\rangle$, while the other
three vanish, so they are in the kernel of the projection. Thus at this
level we have one non-vanishing null vector (a Fock space state which is
both primary and secondary) and three vanishing null vectors. By general
arguments which are well known for the Verma modules of $c=1$ and $c=25$
theories of a single free boson \ref\kato{M. Kato and S. Matsuda, in {``\it
Conformal Field Theory and Solvable Lattice Models''}, Adv. Studies in Pure
Math. 16, Ed. M. Jimbo, T. Miwa and A. Tsuchiya (Kinokuniya,
1988).}\lianetc, a vanishing null vector gets replaced by a Fock space
state which is not a secondary. This can therefore be an extra primary
(which is the situation for $c=1$) or a state which is neither primary nor
secondary (which is the case at $c=25$). In the present case we can easily
find the extra states, which turn out to be the ones created by the fields
$c\del\phi$, $c\del c\del\phi$ and $c\del c c\del^2 c$. All three are
non-primary for the twisted $N=2$ algebra. Thus this situation resembles
more closely the case for $c=25$.

We conclude that the $c=1$ string theory Hilbert space, in the framework
of twisted $N=2$ topological symmetry, consists of the Fock vacua in
Eqs.\vac,\vactwo, along with $N=2$ secondaries (some of which are null),
and extra Fock space states which are neither primary nor secondary (we
will argue below that this is true in general).

Let us now continue with the example above and see where the cohomology of
the BRST charge $G^+_0$ appears. It is well-known that states in the
cohomology appear only for conformal dimension $h=0$. Thus in our example,
we may ask which of the five secondaries above $c_1|0\rangle$ and
$c_0c_1|0\rangle$, together with the three extra Fock space states, lie in
the cohomology. Altogether there are four such states, which are created
by the operators 1, $c\del X$, $c\del c\del X$ and $\sqrt2\del c +
c\del\phi$.  The first two are in the relative cohomology with respect to
$G^-_0=b_0$, and correspond to a Lian-Zuckerman state of ghost number 0,
and a discrete state (the chiral part of the zero-momentum dilaton) of
standard ghost number 1. The other two states are not in the relative
cohomology. It is noteworthy that the one state which is a non-vanishing
null vector of the $N=2$ algebra is the first Lian-Zuckerman state (of
ghost number zero).

One can argue that in fact, all the Lian-Zuckerman states
of ghost number 0 are non-vanishing null vectors of the $N=2$
algebra. More precisely, since these states are
cohomology classes, the claim is that each class has
a representative which is a non-vanishing null vector.
To prove this, start with the chiral operators
\eqn\lz{\eqalign{ x(z) &= \left( cb + {i\over\sqrt2}(\del X
- i\del\phi)\right) e^{{i\over\sqrt2}(X + i\phi)} \cr
y(z) &= \left( cb - {i\over\sqrt2}(\del X + i\del\phi)
\right) e^{-{i\over\sqrt2}(X - i\phi)}\cr}  }
which generate the ground ring of ghost number zero cohomology classes
\wittring . One can check explicitly that the commutator of $L_n$,
$G^\pm_n$ and $J_n$ with these operators vanishes for $n>0$. Now since
operator products of these operators, viewed as cohomology classes, are
non-singular, it follows by repeated commutation that all the chiral
ground ring elements are primaries of the $N=2$ algebra. However, since
they are not Fock vacua, they must appear in the Fock space as secondaries
of $N=2$. Hence they are non-vanishing null vectors.

Thus we have seen, in brief, how the $N=2$ twisted superconformal algebra
organizes the states of $c=1$ string theory. Tachyons are primaries,
discrete states of the same ghost number are secondaries and discrete
states of ghost number zero (ground ring generators) are non-vanishing
null vectors. A more detailed analysis of this system can be done, to
provide the complete classification of $c=1$ string theory physical
states in $N=2$ language, and explore its consequences
\ref\gjm{D. Ghoshal, D. Jatkar and S. Mukhi, work in progress.}.

\newsec{A Special $N=2$ Model}
\subsec{Description of the Coset Model}
In this section we study a special $N=2$ supersymmetric conformal field
theory which we argue, here and in following sections, to be equivalent to
$c=1$ matter coupled to gravity at non-zero cosmological constant.  From
the discussion in Section 2, it is clear that such a superconformal
model should have untwisted central charge $\hat c =3$.  The model is a
supersymmetric Kazama-Suzuki (KS) coset model
\ref\ks{Y. Kazama and H. Suzuki, Nucl. Phys. B321 (1989) 232.}
$$\widehat{SL(2,R)\over U(1)}$$
at level $k=3$
(One can also think of this as $SU(2)$ at level $k=-3$. Here and in the
rest of the paper, we use the conventions that for $SU(2)$, $c=3k/(k+2)$,
while for $SL(2,R)$, $c=3k/(k-2)$.) In this way of writing it we have
hidden the fermionic degrees of freedom.  It is well known that we can
write these in terms of free fermions, which in this case we denote by
$b,c$, and the coset can then be thought of as
\eqn\kscos{{SL(2,R)\times U(1)\over U(1)}}
The computation of the central charge is
$${\hat c}={1\over 3}[{3k\over k-2}+1-1]={3\over 3-2}=3$$
the desired value. The value $k=3$ is the only level for $SL(2,R)$
KS models which gives $\hat c=3$.  The
supersymmetry currents of the $N=2$ algebra
are given by
\eqn\gen{G^+=c J^+\quad\quad G^-=b J^- }
where $J^\pm$ are the raising and lowering operators of $SL(2,R)$.
The $N=2$ $U(1)$ current $J^Q$ is given by
\eqn\uone{J^Q={k\over k-2}cb-{2\over k-2}J^3=3cb-2 J^3}
where $J^3$ is the third generator of $SL(2,R)$.
It is easy to see that with this definition,
$G^\pm$ have charges $\pm 1$ under $J^Q$, as they should.
The $U(1)$ current $J^q$ that we are modding out (i.e., the $U(1)$
in the denominator of \kscos ) is given by
\eqn\umod{J^q=-cb +J^3}
Note that $J^q(z)J^Q(0)\sim 0$ as expected (we
use $J^3(z) J^3(0)\sim -k/2z^2$).
In the untwisted version of this theory, $b$ and $c$ have
spin 1/2.  But as discussed in the previous section,
only twisted $N=2$ theories are related to the bosonic string.
The twisting changes the spin content of the fields by
$$s\rightarrow s-{Q\over 2}$$
where $Q$ is the $N=2$ U(1) charge of the field.  This in
particular means that the spins of the fields change according to
$$c: \qquad {1\over 2}\rightarrow {1\over 2}-{k\over 2(k-2)} =
{1\over 2}-{3\over 2}=-1$$
$$b: \qquad {1\over 2}\rightarrow {1\over 2}+{k\over 2(k-2)} =
{1\over 2}+{3\over 2}=2$$
$$J^+: \qquad 1\rightarrow 1+{1\over k-2}=2$$
$$J^-: \qquad 1\rightarrow 1-{1\over k-2}=0$$
$$G^+: \qquad {3\over 2}-{1\over 2}=1$$
\eqn\spins{G^-: \qquad {3\over 2}+{1\over 2}=2}
{}From the spin content of this theory at the particular value $k=3$ we
see an amazing similarity to bosonic strings. Namely, the $b,c$ have
acquired {\it precisely} the right spins to be identified with the usual
ghosts of bosonic strings, $G^+$ is to be identified with the string BRST
operator, and $G^-$ with the usual $b$ ghost.  There is a slight puzzle in
this correspondence, in that $G^-=b J^-$ contains an extra $J^-$ piece
besides the $b$ ghost. We will return to this point when we discuss the
KPZ version of $c=1$ theory in the next section. From Eq.\spins, $J^-$ has
spin zero. Thus if, as in the KPZ approach, we set it to a constant, we
get the precise correspondence with the usual expression in the DDK
version of $c=1$ string theory, whose $N=2$ structure was discussed in the
previous section.

In order to better understand the connection with the $c=1$ matter theory
coupled to gravity, it is important to find the $Q^+=\int G^+$ cohomology
of the KS model. Those are the states which are to be identified with
physical states of the bosonic theory, since $Q^+$ is the analog of the
BRST operator. In order to do this, it is convenient to go to the Ramond
sector of the theory, and then use spectral flow to write the chiral
fields which form the cohomology of $Q^+$ in terms of fields. This is
just the familiar story that the Hilbert space of the topological theory
is best viewed in the Ramond sector but the operators in the NS sector
\ref\ttstar{S. Cecotti and C. Vafa, Nucl. Phys. B367 (1991) 359.}.  Note
that under spectral flow the $J^3$ charge and $cb$ ghost charge $G$ shift
respectively by
\eqn\specflo{J^3\rightarrow J^3+{3\over 2} \qquad G\rightarrow G+{3\over 2}}
To construct the states of the $G/H$ coset we start with primary states of
the $G$ model, act on them with the $G$ current algebra, and decompose the
result into $H$ representations.  For us $G$ consists of the $b,c$ system
whose representation is just free fermion fock space, and the $SL(2,R)$
Hilbert space. The latter has representations of four types, built on highest
weight states (HWS), lowest weight states (LWS), states which are both,
or states which are neither. A HWS (LWS) is one which is annihilated by
$J^+_0(J^-_0)$:
$$J^+_0|HWS\rangle =0= J^-_0|LWS \rangle $$
A HWS with $J^3=j$ will be denoted $D_j^-$.  Using the $J^-_0$ operator we
get the $J^3$ eigenvalues $j,j-1,j-2,...$, which form an infinite
dimensional representation, unless we encounter a zero vector (which would
then correspond to an LWS).  A LWS with $J^3=j$ will be denoted $D_j^+$.
Using the $J^+_0$ operator we get the $J^3$ eigenvalues given by
$j,j+1,j+2,...\ $.  We will be primarily interested in the case where $2
j\in {\bf Z}$, but will comment on the generalizations below.
Representations which have neither HWS nor LWS turn out to have a trivial
$Q^+$ cohomology, so we will ignore these in the subsequent discussion.
The ones which have both can be viewed as HWS (LWS) representations for
which the $J^-_0$ ($J^+_0$) operator raised to some power annihilates the
state. Thus the raising or lowering terminates at a finite number of steps
and gives rise to a finite dimensional representation of $SL(2,R)$.

To study the $Q^+$ cohomology it is useful to note that
\eqn\susa{\{ Q^+,Q^-\} =L_0}
where $Q^-=G^-_0$ and $L_0$ is the scaling operator in the Ramond sector
(shifted by $-\hat c/8$).  From this it follows that all the cohomology
states of $Q^+$ are represented by zero eigenstates of $L_0$.  To see
this, suppose to the contrary that there is a cohomology state of $Q^+$
given by $|\alpha \rangle$ whose $L_0$ eigenvalue is $h\ne 0$.  Since
$|\alpha \rangle $ is a cohomology state, it must be annihilated by $Q^+$.
Then using \susa\ we deduce that
$$|\alpha \rangle ={1\over h}Q^+(Q^-|\alpha \rangle)=Q^+|\beta \rangle$$
where $|\beta \rangle ={1\over h}Q^-|\alpha\rangle$. It follows that
$|\alpha \rangle$ is cohomologically trivial. So all the non-trivial
cohomology states of $Q^+$ have zero eigenvalues of $L_0$. In a unitary
theory an eigenstate of $L_0$ with eigenvalue $0$ is guaranteed to be
cohomologically non-trivial \ref\lvw{W. Lerche, C. Vafa and, N. Warner,
Nucl. Phys. B324 (1989) 427.}\ but for a non-unitary theory, like the one
we are dealing with, this is no longer true. So we only have a necessary
condition for a non-trivial cohomology element. Note that we can have two
distinct types of cohomology elements depending on how the $Q^-$
cohomology is realized. There are $Q^+$ cohomology states that are
annihilated by both $Q^+$ and $Q^-$, and there are others which form pairs
of the form $|\psi \rangle $ and $Q^-|\psi \rangle $.

Let us see what the condition for vanishing of $L_0$ implies.  Let us
consider a HWS $D_j^-$.  Then the formula for $L_0$ of the coset obtained
from the primary states of the $G$ theory is given by \ks\
\eqn\sdim{L_0={-j(j+1)\over k-2}+{1\over 8}+{q^2\over k-2}-{\hat c \over 8}
=-j(j+1)+q^2-{1\over 4}}
where $q=j\pm {1\over 2}$ is the $U(1)$ charge of $j^q$ and
the sign refers to which of
the spins of the $b,c$ system we take.  Note that if we take
$q=j+{1\over 2}$, i.e. ghost number $-{1\over 2}$
 then $L_0=0$.  In fact it is easy to see that
this is a non-trivial cohomology state,
using the fact that $Q^+=c_0J^+_0+...$.
 It is also annihilated by $Q^-$.  This gives
rise, using spectral flow, to a cohomology state with
ghost number
$$G=-{1\over 2}+{3\over 2}=1$$
according to equation \specflo .

Now we search for other states with $L_0=0$.  We will focus first on
states in the coset with $L_0 \leq 0$, because the states with $L_0=0$ can
be built on top of such `ground states'.  In a HWS representation, the
envelope of states in the $G$ theory can be obtained by either applying
$(J^+_{-1})^r$ or $(J^-_0)^s$ on the highest weight state (see
\tfig\FiglabelA ).
Let us consider first the $L_0$ of states obtained by acting
$(J^+_{-1})^r$ on the HWS.  It will increase the energy of the $G$ theory
by $r$ units and it will shift $q\rightarrow q+r$.  Therefore the change
in $L_0$, starting from $L_0=0$ is
\eqn\chl{\Delta L_0=L_0=r+(q+r)^2-q^2=r(2q+1)+r^2=r((2j+2)+r)}
Since $r\geq 0$ the condition that this gives $L_0\leq 0$ is
that
\eqn\rcond{j\leq -1 \qquad {\rm and}\qquad 0\leq r \leq -2j-2}
For these values of $j$ and $r$, the ground state of the coset has
negative dimension.  By using the other raising operators we can obtain
states with $L_0=0$ only for this range of $j$ and $r$.  Considering now
the $(J^-_0)^s$ action, we get
\eqn\achl{\Delta L_0=L_0=(q-s)^2-q^2=-2qs+s^2=s(s-2j-1)}
and the condition for this $L_0$ to be negative is that
\eqn\scond{j\geq {-1\over 2} \qquad {\rm and} \qquad 0\leq s \leq 2j+1}
We thus see that for a fixed $j$, we have only a finite number of
potential candidates for the cohomology states of $Q^+$, because there are
only finitely many combinations of raising operators that can raise the
energy to $L_0=0$.

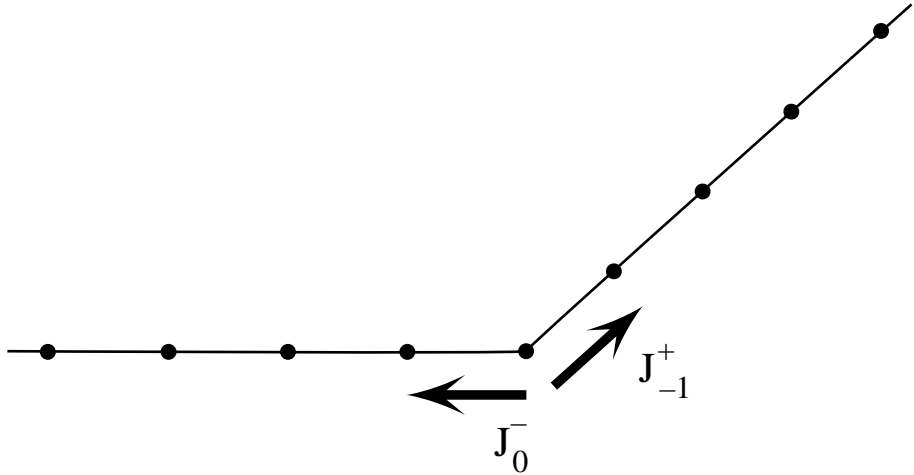
\ifigure\FiglabelA{\baselineskip=12pt The envelope of states in a HWS
representation is
obtained by acting with $J_{-1}^+$ or $J_0^-$ on the highest state.}{Figo}{2.5}

We have carried out this discussion for a HWS $D_j^-$.  By a CPT
transformation a similar story holds for a LWS $D_{-j}^+$, where we have
to flip the signs of all charges in the process.  In particular the LWS
itself will correspond to a non-trivial cohomology element with ghost
number $G=2$.  Similarly using $(J^-_{-1})^r$ and $(J^+_0)^s$ we get a
finite range of values of $j,r$ and $j,s$ with negative $L_0$ just as in
\rcond\ and \scond .

To proceed further we have to discuss which representation of affine
$SL(2,R)$ Kac-Moody algebra we should use. There are 4 natural choices: the
irreducible representation $I$, which is natural for $k<0$ (which can be
thought of as $SU(2),~k>0$), the Wakimoto representation $W$, the dual
Wakimoto representation $W^*$ and the Verma module $V$. The one which
turns out to be most clearly related to $c=1$ theory coupled to gravity is
$W$, though there are some intriguing features for the other
representations which we mention later.

To begin with we recall how the Wakimoto fields
realize the $SL(2,R)$ algebra.  There are three bosonic fields,
$\beta, \gamma $ and $\phi $. The $SL(2,R)$ currents are given by
$$J^+=\beta \gamma^2-{\sqrt 2}\gamma \partial \phi +3 \partial \gamma$$
$$J^3=\beta \gamma -{1\over \sqrt 2}\partial \phi$$
\eqn\waki{J^-=\beta }
In the untwisted theory $\beta$ has dimension $1$ and $\gamma$ has
dimension zero, but in the twisted theory which is the one we are
interested in, it is easy to see that $\beta$ will have spin 0, and
$\gamma$ spin 1.  They satisfy the OPE $\gamma (z) \beta (0)
\sim -1/z$. $\phi$ is a scalar field with a background charge.
In the twisted theory the background charge of $\phi$ is exactly
the same as that of the Liouville field in $c=1$ theory, and we
will later identify it with that.

So far we have a good description of the field content of the $SL(2,R)$
theory and the fermionic $b,c$ system. Now we need an equally convenient
representation for the coset model. This is readily obtained by gauging
the $J^q$ current. As is familiar \ref\gawed{K. Gawedzki and A.
Kupiainen, Nucl. Phys. B320 (1989) 625.}, this leads to the introduction
of a scalar field $X$, whose derivatives are related to the gauge field,
and a fermionic ghost system $B,C$ of dimension $1,0$.  These ghosts can
be used to define the BRST operator for the $U(1)$ gauge symmetry:
\eqn\uonec{Q_{U(1)}=C(J^3-cb -i{\partial X\over {\sqrt 2}})=C(
\beta \gamma -cb -\partial X^- )}
where
\eqn\xphi{X^{\pm}={\mp iX+\phi \over \sqrt 2}}
Note that the $X$-field is a $c=1$ system whose radius is fixed by the
level of the $j^q$, which is one, to be the {\it self-dual} value. Let us
summarize what we have found as the Hilbert space of the coset:
$${\cal H}=[\phi ]+[X]+[b,c]+[\beta, \gamma]+[C,B]$$
Note that this differs from the standard DDK Hilbert space of $c=1$ {\it
only} by the appearance of bosonic ($\beta, \gamma$) and fermionic ($C,B$)
fields of spins (0,1).  In order to make contact with KS theory we have to
reduce the Hilbert space of this theory by first imposing the $U(1)$
constraint via the BRST operator $Q_{U(1)}$:
$${\cal H}_{KS}={\cal H}_{Q_{U(1)}}$$
Then we can consider the cohomologies of the KS theory
on this reduced Hilbert space by acting with
$$G^+= c J^+= c(\beta \gamma^2 -{\sqrt 2}\gamma \partial \phi
+3 \partial \gamma )$$
In this representation $G^-$ is particularly simple:
\eqn\cgm{G^-=b\beta }
Note that using the constraint \uonec,
the $N=2$ current $J^Q$ \uone\ can be written as
$$J^Q=cb-i{\sqrt 2} \partial X$$
This is precisely what is expected for $c=1$ theory based on our analysis
in section 2 (see \topalg\ with $\eta =X$ and $y=i \sqrt 2$ ).

The cohomology spectrum of this KS theory is analyzed by E. Frenkel in
appendix A, using techniques developed in \ref\frenko{E. Frenkel, Phys.
Lett. B 286 (1992) 71.}.
Let us consider a HWS $D_j^-$ of Wakimoto. His results are
as follows: Apart from the cohomology we discussed before, there is
precisely one pair of cohomology states of ghost number $(0,1)$ for the
range of $r$ and $j$ given by \rcond\ (with $q=j+r+1/2$) and one pair of
cohomology states of ghost number $(1,2)$ for the range of $r$ and $j$
given by \scond\ (with $q=j-s+1/2$) (see \tfig\FiglabelB). Moreover, as
we will see explicitly below, the higher ghost number cohomology state
gets mapped by $Q^-$ to the lower one. This is not surprising in view of
the fact that $Q^-$ cohomology, being quadratic in fields, is non-trivial
only for the HWS itself. For the LWS $D_{-j}^+$ apart from the ghost
number 2 cohomology state that we mentioned before there is a pair of
cohomology elements with $r$ and $j$ satisfying condition \rcond\ (with
$q=-j-r-1/2$) with ghost numbers $(2,3)$ and the ones satisfying condition
\scond\ (with $q=-j+s-1/2$) with ghost numbers $(1,2)$.  In both classes
the highest ghost number state is mapped to the lower one by the action of
$Q^-$, as discussed before.

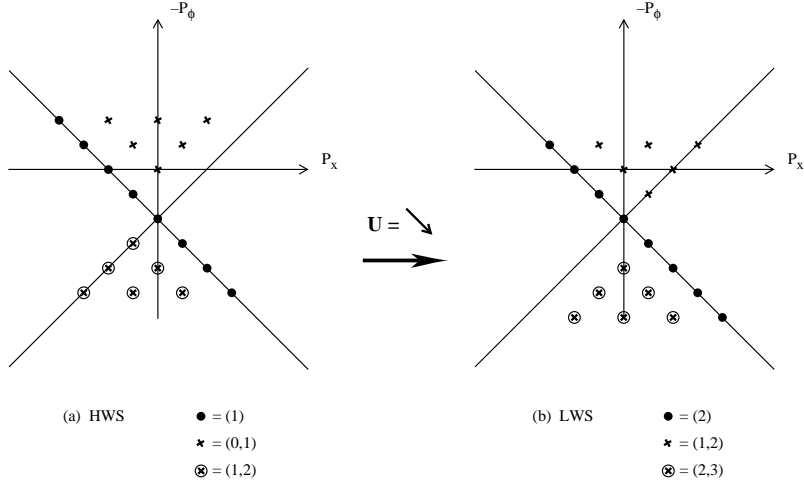
\ifigure\FiglabelB{\baselineskip=12pt The cohomology spectrum for HWS (a) and
LWS (b).
The spectrum of LWS can be obtained from that of HWS by operating
with $U$ which shifts the matter and Liouville momenta as shown, and
increases ghost number by one.  The degeneracy and ghost numbers of the
cohomology states are represented in the figure.
A given representation leads to states in the same horizontal
line.  The cosmological constant operator in (a) corresponds
to $D^-_{-1/2}$ with $j$ increasing as we go down in the figure.
The cosmological constant state in (b) corresponds
to $D^+_{1/2}$ with $j$ increasing as we go up in the figure.
Note that the positive Liouville momentum in our convention
corresponds to the lower part of the figures.
}{Figt}{2.5}

In the next two subsections, we will explicitly construct representatives
of the cohomology of this coset model, and display their similarity to
those of the $c=1$ string.

\subsec{Explicit Construction of the Cohomology: Highest Weight States}
Let us first start with the Hilbert space generated by HWS.  The simplest
cohomology states, as we discussed before, arise from the highest weights
themselves. The cohomology state considered below Eq.\sdim (coming
from $D_j^-$ with $q=j+1/2$) corresponds to the field
\eqn\tachhws{V_j=c\ {\rm exp}{\sqrt{2} [(j+{1\over 2}) iX + (j+{3\over
2})\phi ]}
=c\ {\rm exp}[{(2j+1) X^- +\sqrt{2} \phi}]}
To see this, note that the exponent of $\phi$ is $j+3/2$ because the
spectral flow shifts the $J^3$ eigenvalue by $3/2$.  Also, we started with
a ghost number $-1/2$ state and spectral flow shifts that by $3/2$ to
ghost number $1$, thus explaining the appearance of $c$ in the above. The
highest weight vacuum state is defined as the state annihilated by
$\gamma_0$ and all higher modes, and $\beta_1$ and all higher modes. Since
in the twisted theory $\gamma$ and $\beta$ have spins $1,0$ respectively,
this is the vacuum given by the path integral, and requires no further
insertions.  The above expression for these cohomology states are {\it
precisely} the same expressions as appear in $c=1$ DDK theory for tachyon
states of ghost number 1. The above tachyon states for all $j$ have a
particular chirality (from the target space point of view), as they depend
on $X^-$ and not on $X^+$.

To make the contact with DDK theory more explicit, let us re-label these
states as follows. For $j\le -1/2$, define $s=-(j+1/2)$ and denote the
corresponding states
\eqn\ypsms{{\tilde Y}^+_{s, -s} = c e^{-\sqrt2 isX}
e^{\sqrt2(1-s)\phi},\qquad s\ge 0}
For $j\ge -1/2$, define $s=j+1/2$ and denote the state
\eqn\ymss{{\tilde Y}^-_{s, s} = c e^{\sqrt2 isX}
e^{\sqrt2(1+s)\phi},\qquad s\ge 0}
These states can now be compared with those of Ref.\wittring\
corresponding to the same labels with the tildes dropped.

How do we find the rest of the states?  One series of states is easy to
find. From the cohomology computation in the appendix, it is clear that
one state of ghost number 1 arises with the $X$ momentum exactly that
allowed by the boundary of the conditions \rcond\ and \scond. Since there
is exactly one state in our theory with that $X$ momentum and with ghost
number 1 and dimension zero, it is straightforward to identify it.  For
HWS with $j\leq -1$, we set $r=-2j-2$, which is the boundary of Eq.\rcond.
The state is obtained by acting with $(J^+_{-1})^{2|j|-2}$ on the HWS.
Using \waki, the action of this operator on the HWS is equivalent to the
action of $\gamma_{-1}^{2|j|-2}$. So the state coming from the boundary of
\rcond\ with $j\leq -1$ is
\eqn\rbound{B_j=c\ \gamma^{2|j|-2} {\rm exp}(\sqrt{2}(j+{3\over 2})\phi )
+(i\sqrt{2}(-j-{3\over 2})X)=c \
\gamma^{2|j|-2}{\rm exp}((2j+3)X^+)}
where we have used the $U(1)$ constraint \uone\ to find the $X$ momentum,
and also the fact that in the twisted theory $\gamma$ has dimension 1.  It
is easy to check directly that these fields commute with $Q^+$ and are not
$Q^+$ trivial. They correspond to ghost number 1 discrete states of $c=1$
closest to the positive $X$ momentum and negative Liouville momentum (see
fig. 2).

Again it is convenient to re-label these states to display the
correspondence with DDK theory. The case $j=-1$ has already been
counted in Eq.\ypsms, so take $j\le -3/2$ and let $s=-(j+3/2)$. Then,
define
\eqn\aoss{{\tilde a}{\tilde {\cal O}}_{s,s} = c \gamma^{2s+1}
e^{-2sX^+}}
The notation and the precise correspondence of these states with
Ref.\wittring\ will become clear shortly.

As discussed before, the $Q^-$ cohomology is trivial on these states,
so we can obtain another set of $Q^+$ cohomology states just by acting on
these with $Q^-=G^-_0$. This is given by the residue of the double pole of
the OPE of $b\beta$ with the above, which is (for $j< -1$)
\eqn\discr{B'_j= \gamma^{2|j|-3}{\rm exp}((2j+3)X^+)}
With the same definition of $s$ above, these states may be re-labelled
\eqn\oss{{\tilde {\cal O}}_{s,s} = \gamma^{2s}
e^{-2sX^+}}
They have ghost number zero and should generate a subset of the ground
ring. In particular we see that the identity operator is already in the
list, corresponding to $j={-3/2}$. The state which has the right momentum
to correspond to the ground ring element $x$ of $c=1$ theory \wittring\
comes from $j=-2$:
\eqn\xgen{{\tilde x}={\tilde{\cal O}}_{\half,\half}=
\gamma {\rm exp}(-X^+)}
Moreover all the rest of the above ghost number zero states can be written
as powers of this element:
$$B'_j= \gamma^{2|j|-3}{\rm exp}((2j+3)X^+)={\tilde x}^{2|j|-3}$$
corresponding to
$${\tilde{\cal O}}_{s,s} = {\tilde x}^{2s}$$
Thus the ring structure is exactly the same as that generated by $x$ in
$c=1$ DDK theory.

A natural question is whether $\tilde x$ and $x$ are related. Indeed they
are: on modifying the $\tilde x$ generator by adding $Q_{U(1)}$
trivial fields, we get
$$\tilde x \sim \gamma {\rm exp}(-X^+)-
\{ Q_{U(1)},B \beta ^{-1}{\rm exp}(-X^+)
\}=\beta^{-1}(cb+\partial X^-) {\rm exp}(-X^+)=$$
$$=\beta^{-1}x$$
where we have been somewhat formal in taking negative powers of $\beta$.
Since $\beta$ has dimension 0 this is not such an unnatural thing to do.
We will make this more precise below when we bosonize the $\beta,\gamma$
system. At any rate, on the Hilbert space omitting the kernel of $\beta$
we see that the two are related.

Now we come to the question of how to find the other discrete states
coming from HWS. In the $c=1$ theory they can be obtained by acting with
the $SU(2)$ raising and lowering operators which are present at the
self-dual radius. There is a similar symmetry in our case, which follows
from an application of the ideas in \wittring .  The idea in the context
of bosonic strings is to generate a $(1,0)$ current from the $(0,0)$
cohomology states by acting with $b_{-1}$. The analog of this operator for
us is $G^-_{-1}$, which means we consider the OPE of $b\beta $ with the
fields and take the residue of the single pole term.  Applying this to the
tachyon state $V_{-3/2}$ (which in $c=1$ gives rise to the lowering
current ${\rm exp}(-i\sqrt{2}X)$) we obtain
\eqn\deflo{K^-=\oint b
\beta \cdot (c \quad {\rm exp}(-i \sqrt{2} X))=
\beta {\rm exp}(-i \sqrt{2} X)}
The minus superscript over $K^-$ is to remind us that the corresponding
charge decreases the $X$ momentum by one unit. The operator $K^-$ is the
analog of the lowering operator of the $c=1$ theory, but it contains an
extra factor of $\beta$. Using $K^-$ we can generate all the discrete
states with ghost number 0 (and the corresponding ghost number 1 states
which can be mapped to them by $Q^-$). We apply $K^-$ up to $2s$ times on
the states corresponding to ${\tilde{\cal O}}_{s,s}$ The states so
obtained are denoted
\eqn\osn{{\tilde{\cal O}}_{s,n} = (K^-)^{s-n} {\tilde{\cal O}}_{s,s}}

As an example, let us consider $K^-$ acting on $\tilde x$ to obtain the
analog of the $y$ generator. We find that it is given by
$${\tilde y}=\int \beta {\rm exp}(-i \sqrt{2} X)(\gamma {\rm exp}(-X^+)=
(\beta \gamma -i\sqrt{2}\partial X){\rm exp}(-X^-)$$
Using the $U(1)$ constraint, this is easily seen to be equivalent to
$${\tilde y}=(cb+\partial X^+){\rm exp}(-X^-)=y$$
which is exactly the same expression as in the DDK theory.  We could have
obtained this answer equivalently by applying $K^-$ on $\beta^{-1}x$,
which would reduce the computation to exactly the one which gives rise to
$y$ in the usual $c=1$ theory (as $\beta$ in $K^-$ plays no role in the
OPE, and just cancels the factor of $\beta^{-1}$). The subspace of the
chiral ring generated by $[{\tilde x},{\tilde y}]$ is thus isomorphic to
the ring generated by $[\beta^{-1}x,y]$, which is in turn the same as the
$c=1$ ring $C[x,y]$ because $x$ and $y$ have no $\beta$ excitations in
them. We thus obtain the exactly the same expression as in DDK theory for
{\it all} discrete states of ghost number 0, modulo factors of
$\beta^{-1}$ which accompany each factor of $x$.  As we discussed, the
ghost number 1 analogs of these states can be obtained by applying $K^-$
to $D_j$ a number of times.

There is another way of doing this, which makes the connection
with $c=1$ DDK theory more transparent. Note that the ghost
number one state paired with the identity is
$$D_{-3/2}=c\gamma $$
In the DDK theory this is given by an operator $a$
\wittring\ defined by
$$a=\{ Q,{\phi \over {\sqrt 2}}\}$$
It turns out that if we use our expression for $Q^+$, we find
$${\tilde a}=\{ Q^+,{\phi \over {\sqrt 2}}\}=c \gamma$$
Note that ${\tilde a}^2=0$ and $a$ commutes with $Q^+$ but not
with $Q^-$:
$$\{ Q^-,c\gamma \} =1$$
This implies that we can obtain the pairs of discrete states simply by
multiplying the one of lower ghost number with ${\tilde a}=c\gamma$, which
increases the ghost number by 1. This explains the notation in Eq.\aoss.

So far we have discussed the discrete states coming from HWS with $j\leq
-1/2$.  For HWS with $j\geq 0$ we can again use the tachyon states $V_j$
and act on them with $K^-$ a number of times to find all the discrete
states with ghost number 1. Since they are obtained from the tachyon
state $V_j$ which is annihilated by $Q^-$, and $K^-$ commutes with $Q^-$,
it follows that $Q^-$ also annihilates all these states. So the pair which
they form must be with a ghost number 2 discrete state, obtained by acting
on them with $\tilde a$, as discussed before. Thus we have
\eqn\ymsn{{\tilde Y}^-_{s, n} = (K^-)^{s-n} c e^{\sqrt2 isX}
e^{\sqrt2(1+s)\phi},\qquad -s\le n \le s}
and their partners ${\tilde a}{\tilde Y}^-_{s, n}$, for which, however, the
range is $-s\le n \le s-1$. The reason is that ${\tilde a}$ has a vanishing
OPE with ${\tilde Y}^-_{s,s}$.

Note that the state with the maximum number of $K^-$ operators acting on
$V_j$ is the one which saturates \scond. This, however, corresponds to a
tachyon state with the opposite chirality. Noting that $J_0^-$ is
represented by $\beta$, and taking into account \uonec, we find that this
tachyon is given by
$${\tilde V}_j=\beta^{2j+1}\ c\ {\rm exp}[\sqrt{2}(-i(j+{
1\over 2})X+(j+{3\over 2})\phi)]$$
$$=\beta^{2j+1}\ c\ {\rm exp}[(2j+1)X^+ +{\sqrt 2}\phi]$$
where $j\geq 0$. In the notation of Eq.\ymsn, this is just
${\tilde Y}^-_{s,-s}$, which has the same expression as in
$c=1$ DDK theory except
for some extra $\beta$ dependence.
Note that we can similarly obtain the tachyon state with positive
chirality of ghost number 2 by
acting with $\tilde a$ which gives $\beta^{2j}~c\partial c
{}~{\rm exp}[(2j+1)X^++{\sqrt 2}\phi ]$.

We have thus finished discussing how we can explicitly obtain all the
cohomologies coming from HWS depicted in fig. 2 in the Wakimoto
representation. Note that we have obtained all the tachyon states (of
both ghost numbers) with positive Liouville dressing and positive
chirality, and half of the tachyon states with negative chirality (those
with ghost number 1). Of the discrete states, we have found pairs with
ghost numbers (0,1) and negative Liouville dressing, and pairs with ghost
numbers (1,2) and positive Liouville dressing (see fig. 2). These results
may be summarized as follows:
\eqn\hwsstate{
\eqalign{&(i)\qquad  Y^+_{s, -s} = c\ {\rm exp}({- i \sqrt2 s X+
\sqrt2(1-s)\phi})\qquad   ( s  \ge 0. )\cr
&(ii)\qquad Y^-_{s,n} = (K^-)^{s-n} \cdot c\
{\rm exp}({i\sqrt2 s X+\sqrt2 (1+s) \phi})\qquad   ( - s \le n \le s )\cr
&(iii)\qquad {\tilde a} Y^-_{s,n},\qquad  ( -s \le n \le s-1)\cr
&(iv)\qquad O_{s,n}=\beta^{-(s+n)}x^{(s+n)}y^{s-n}\qquad (s \geq 0 \quad ,
-s \leq n \leq s)\cr
&(v)\qquad  {\tilde a} O_{s,n}  \qquad (s \geq 0 \quad ,
-s \leq n \leq s)\cr}   }
where $x,y$ are as in the $c=1$ DDK theory,  $s$ and $n$ are both integer
or half-integer, $K^- = \int \beta e^{- \sqrt2 i X}$,
and $\tilde a =c \gamma$.

The rest of the states come from LWS representations, to which we now turn.

\subsec{Explicit Construction of the Cohomology: Lowest Weight States}
In obtaining the cohomology states from LWS we can use the same ideas as
above.  However, it turns out that there is an even faster way of getting
all these states:  The HWS and LWS representations turn out to be
equivalent in the coset language. This follows from the fact that if we
have a coset of the form $G/H$ the common center of $G$ and $H$ generate
an isomorphism between coset characters \ref\ms{G. Moore and N. Seiberg,
Phys. Lett. B220 (1989) 422.}\lvw. In the case at hand the action on
$SL(2,R)$ is to take a HWS of spin $j$ and map it to LWS of spin
$j+(k/2)$. For us, this means that
$$D_{j}^- \rightarrow D_{j+{3\over 2}}^+$$
In addition, the ghost number and the $U(1)$ charge are increased by one
unit. In the Wakimoto realization, this is the spectral flow by the
operator
\eqn\spwa{U=c\delta (\beta ){\rm exp}(X^-)}
where $\delta (\beta )$ (which also contributes $+1$ to the $U(1)$
constraint) is needed to change the HWS to a LWS (annihilated by $J^-$).
The operator $U$ can be viewed as the spectral flow generated by the
constraint current \uonec .  From that viewpoint, it may be natural to
assume that we have to identify the two representations and that they
should be viewed as picture-changed versions of one another. Taking into
account the shift in the ghost number and the $X$ and $\phi$ momenta
obtained by acting on HWS with $U$, we immediately get the spectrum of
cohomologies depicted in fig. 2. Note in particular that if we put the two
spectra together {\it we end up with precisely the same  degeneracy of
cohomology classes as in the} $c=1$ {\it DDK theory} if we do not identify
the two representations!  We will have more discussion on this below.

Just to give a sample of the LWS that we obtain, we
write the tachyon states coming from the LWS representation
$D_j^+$ which are of ghost number 2:
$$T_j=c\partial c \ {\rm exp}[(2j-1)X^-+{\sqrt 2}\phi ]\quad
\delta (\beta )$$
We can also obtain the ghost number 3 partners of these states
by multiplying by ${\tilde a}=c\gamma $ as was the case for HWS.

We also write explicit expressions for tachyon states with negative
Liouville dressing and opposite chirality, which are obtained from LWS
with $j\leq 0$ (these can be obtained by multiplying ${\tilde x}^n$ by $U$
or by acting on $T_j$ (with $j\leq 0$) with $(J^+_{0})^{2|j|+1}$, which on
LWS is equivalent to acting with $\gamma_0^{2|j|+1}$):
$${\tilde T}_j=c\partial c \ {\rm exp}[(2j-1)X^+ +\sqrt{2} \phi ]
\delta^{2|j|+1}(\beta )$$
where the superscript over $\delta$ denotes the order of derivative of the
delta function. We can use $Q^-$, which does not annihilate these states,
to obtain the corresponding ghost number 1 tachyon states. We can now
obtain all the ghost number 1 and 2 discrete states by considering
$U{\tilde x}^n{\tilde y}^m$ and $c\gamma U {\tilde x}^n{\tilde y}^m$.

As an example, let us consider the case $j=-{1/2}$, and write the
corresponding discrete states. In this case there is only one pair of
discrete states, and they are found by the above method to be
$$R=i\sqrt{2}c\delta (\beta )\partial X +c\partial \beta
\delta '(\beta )\qquad {\rm and}\qquad
{\tilde a} R $$
Note that the state $R$ is the analog of the discrete state in the
ordinary $c=1$ DDK theory at zero cosmological constant given by
$c\partial X$.

If we view the coset model as arising from a gauged WZW model, then the
natural question is whether we can realize the Wakimoto fields in terms of
the field variables of the WZW theory. It was shown in
\ref\ger{A. Gerasimov, A. Morozov, M. Olshanetsky and A. Marshakov, Int.
J. Mod. Phys. A5 (1990) 2495.}\ that this is indeed possible. There is
one interesting subtlety: The Wakimoto fields have one extra degree of
freedom relative to the fields of the WZW theory, and this leads to the
appearance of a screening charge given by
\eqn\scre{Q_F=\oint \beta {\rm exp}(\sqrt{2} \phi )}
(and similarly for the right-movers). The existence of this screening
current was seen to be crucial in computations in WZW theory \ger\ in the
Wakimoto realization (and in particular needed to balance the $\phi$
charge violation). The cosmological constant operator in the ghost number
1 representation is given by $V_{-1/2}=c\ {\rm exp}\sqrt{2}\phi$. When
acted on by $G^-_{-1}{\bar G}^-_{-1}$, which is needed in order to add it
to the action, gives us the current
\eqn\cosmo{G^-_{-1}{\bar G}^-_{-1}V_{-1/2}=
\beta {\bar \beta}{\rm exp}( \sqrt{2} \phi ),}
Thus, the existence of the above screening charge insertion is directly
related to the fact that the cosmological constant is non-zero  for us,
i.e., {\it the screening current of the theory automatically puts
cosmological constant operator insertions in correlation computations in
order to achieve} $\phi$ {\it momentum conservation}. Moreover the
identification of the above operator with the cosmological constant
operator of $c=1$ DDK theory suggests that if we set $\beta$, which is a
dimension zero operator, to a constant:
\eqn\setbet{\beta ={\bar \beta}={\sqrt \mu}}
(where $\mu$ is the cosmological constant) then we should obtain an exact
correspondence with $c=1$ DDK theory. We will see in the next section
that \setbet\ is the precise condition dictated by the KPZ formulation.
Again, we see that setting $\beta$ to a constant makes the correspondence
with $c=1$ theory exact. Put differently, if we {\it count} the number of
times we use the screening current, and if each time we use the (left and
right) currents we put a factor of $\mu$, we get the correct $\mu$
dependence. As it stands the KS theory counts each insertion of screening
operator with weight one and should be viewed as having $\mu =1$.

Even though setting $\beta$ to a constant appears to be relevant for
comparison to the DDK theory, at first sight there seems to be an
obstacle:  The LWS all involve $\delta(\beta )$, and therefore setting
$\beta =\sqrt{\mu }$ will disallow all these states. Thus we will miss
half of our states. Moreover, ignoring the LWS means that we are getting
rid of the dual fields of HWS. In particular this would imply that the two
point functions which pair a HWS with a LWS do not exist, which does not
make much sense.

However, it has been observed \ref\mardot{A.V. Marshakov, Phys. Lett. B224
(1989) 141\semi Vl.S. Dotsenko, Nucl. Phys. B338 (1990) 747.}\ that dual
fields to HWS in the Wakimoto representation can be constructed without
the use of LWS {\it if} we allow negative powers of some fields. In our
case this can be done if we allow negative powers of $\beta$. In fact at
the level of fields, rather than states in the Hilbert space, it makes
sense to talk about the cohomology of chiral fields. Allowing negative
powers of $\beta$ gives rise to new cohomology fields which are in
one-to-one correspondence with those of LWS, and can be obtained from them
by the simple substitution
\eqn\subim{\delta (\beta )\rightarrow \beta^{-1}}
It is easy to see that these are in the cohomology and satisfy the \uone\
constraint. The fields thus obtained will not correspond to states in the
Hilbert space with well defined $SL(2,R)$ quantum numbers, but
nevertheless they are natural for us, as they were in the computations of
$SL(2,R)$ correlation functions in \mardot.

In order to make this a little more precise, it turns out to be useful to
bosonize the $\beta,\gamma $ system \ref\fms{D. Friedan, E. Martinec and
S. Shenker, Nucl. Phys. B271 (1986) 93.}\ in terms of
two scalar fields $u,v$ by
$$\beta ={\rm exp}(u-iv)\qquad \gamma =-i\partial v \ {\rm exp}(-u+iv)$$
where $u,v$ have background charges
(contributing to the energy momentum tensor in the form
$(-\partial^2 u+ i\partial^2 v)/2$). In this form, the
inverse powers of $\beta$ make sense:
$$\beta^{-1}={\rm exp}(-u+iv)$$
Noting that we need an extra $\beta^{-1}$ to absorb the $u,v$ background
charge, it is an easy check that for instance $V_j$ and $T_j$ (with the
substitution \subim) give a non-zero two point function, as do $\tilde
V_j$ and $\tilde T_j$.

With the substitution \subim\ and after setting $\beta $ to a constant
(which we discuss in the next section), we see that {\it all the states of
the} KS {\it model are identically represented as in the } $c=1$ {\it DDK
theory}.  This completes the correspondence with $c=1$ states\foot {The
discussion of cohomology states was only chiral (for left-movers). As the
story for combining the left and right movers is similar to c=1, we did
not bother to discuss this point. In particular the neutrality of the
Hilbert space under $G_0^- -{\bar G}_0^-$ implies that only ${\tilde
a}+\bar{\tilde a}$ can appear when we combine left and right movers, which
is the same thing that happens in $c=1$ DDK theory \wittring.}

Returning to the construction of the dual states to the HWS, we can simply
act on the HWS states in Eq.\hwsstate\ with U given by \spwa, and make the
replacement \subim. This is equivalent to multiplying by $c~\beta^{-1}
{\rm exp}(X^-)$ and taking the regular term in the OPE (this is also true
for the usual $c=1$ DDK theory except for the extra factor of $\beta
^{-1}$). Some of the fields so obtained are:
\eqn\lwsr{
\eqalign{
&(i)\qquad aY^-_{s,s} =
c \del c \ \beta^{-1}\ {\rm exp}(\sqrt2 i s X +\sqrt2
(1+s)\phi )\qquad  ( s \ge 0 )\cr
&(ii)\qquad  aY^+_{s,n} = (K^-)^{s-n}\cdot c \del c\ \beta^{-2s-1}\
{\rm exp}({\sqrt2 i s X+\sqrt2(1-s)\phi}\ ( -s \le n \le s )\cr
&(iii)\qquad  Y^+_{s,n} = (K^-)^{s-n}\cdot c \ \beta^{-2s}
\ {\rm exp}( \sqrt2 i s X+\sqrt2(1-s)\phi )\qquad ( -s \le n \le s )\cr }}
and the rest can be constructed similarly.

We thus see explicitly that except for extra factors of $\beta$ (which go
away if we set $\beta =1$), the states which are in the cohomology of
$Q^+$ (relative to $Q^-$) have exactly the same expressions as in the
$c=1$ DDK theory.

Discrete states of ghost number 1 (which are annihilated by $b_0$) were
used in \wittring\ to construct an area-preserving symmetry algebra for
the ring generated by $x$ and $y$ for the $c=1$ model at zero cosmological
constant. For us, the corresponding object would be obtained by acting
with $G^-_{-1}$ on the ghost number 1 discrete states coming from LWS,
with the substitution \subim, and on the tachyon states $V_j$ from the
HWS. It is easy to see that their action on the ground ring is given by
area-preserving diffeomorphisms exactly as for the usual $c=1$
case\wittring\pol. This is obvious in view of the fact that the only
difference here with respect to the $c=1$ case is some extra $\beta$
dependence, which turns out not to have any effect on the algebra.

For example, the $SU(2)$ subalgebra of the area-preserving symmetry is
represented here by
$$W^+_{1,1} = - \beta^{-1}\ {\rm exp}({\sqrt2 i X})$$
$$W^+_{1,0} = - \beta^{-1} \del\beta + \sqrt2 i \del X $$
$$W^+_{1,-1} = \beta {\rm exp}({-\sqrt2 i X})$$
Note in particular that the current responsible for the change of radius,
$W^+_{1,0}$, now has an extra $\beta$-dependent piece.

This completes the construction of explicit representatives
of the cohomology in the Wakimoto representation.

\subsec{More About the Coset Model}
Let us briefly mention what happens if we consider instead of the Wakimoto
representation $W$, other representations of affine $SL(2,R)$. We first
consider the dual Wakimoto representation $W^*$, which turns out to be
rather interesting.  Choosing $W^*$ means that we choose the same fields
as before but interchange $+ \leftrightarrow -$ which in particular means
that $\beta =J^+$ (and some signs are changed in \waki). This means that
now the spin of $\beta$ is $2$ and that of $\gamma$ is $-1$. Moreover
the field $\phi$ in this representation has {\it no background charge}!

So the spectrum for this theory will consist of two scalar fields $X$ and
$\phi$ with no background charge, two sets of ghost fields $(b,c)$ and
$(\beta,\gamma)$ of spins $(2,-1)$, one fermionic and one bosonic, and a
pair of fermions of spin $(0,1)$. Amazingly, {\it this is the same field
content as topological gravity coupled to} ${\hat c}=1$ {\it topological
matter}!  To make this a little more transparent let us define a complex
scalar field coordinatizing an infinite cylinder
$$Z={(\phi + i X)\over \sqrt{2}}$$
and redefine the fermionic fields $B,C$ by
$$\psi^* =C \quad \psi =B$$
In addition we have two BRST charges $Q^+$ and $Q_{U(1)}$
which now take the form
\eqn\topbr{Q^+=c\beta  \qquad Q_{U(1)}=-\psi^* (\partial Z +cb -\beta \gamma)}
Solving the corresponding cohomology in this case is much easier than in
the case of $W$ because the fields appear only quadratically in the BRST
charges. In particular considering the $Q^+$ cohomology means eliminating
$b,c,\beta $ and $\gamma $, and so $Q_{U(1)}$ reduces to
$$\psi^* \partial Z$$
whose cohomologies are of the form
$${\rm exp}(n Z)$$
for integer $n$. (The fact that $n$ should be integer follows from the
radius of the $X$ field). Precisely this topological theory was
conjectured to be related to the $c=1$ bosonic matter theory in
\ref\Hor{P. Horava, Nucl. Phys. B386 (1992) 383.}. The spectrum of the
theory in this language seems to be smaller than what is required for
$c=1$. However, it is conceivable that the rest of the discrete states in
this set up correspond to gravitational descendants. It would be extremely
interesting to develop this further.

If we work with the irreducible representation $I$ of $SL(2,R)$, obtained by
deleting the null states\foot{This is not necessarily a desirable thing to
do for $k>0$, which is like a negative level for $SU(2)$.}, it turns out
that we get exactly the same states as before except that the boundary
states for HWS with ghost number 1 and for LWS with ghost number 2 will be
missing (see appendix A). If we work with Verma module $V$ of the current
algebra, then for HWS the cohomology for $j< -1/2$ matches that of the $W^*$
module and for $j\geq -1/2$ it matches that of $W$, which is a possibly
interesting mixture of these two cases (see appendix A).

Before concluding this section let us note that it is relatively easy to
change the radius from the self-dual value to an integer $n$ times the
self-dual one. One simply considers an $n$-fold cover of the $SL(2,R)$
theory, which means that the allowed $j$'s are multiples of $1/2n$.  It is
amusing that precisely for these values of radii the existence of a
Penner-like model was conjectured in \DV\ and found in \ref\chic{S.
Chaudhuri, H. Dykstra and J. Lykken, Mod. Phys. Lett. A6 (1991) 1665.}
(See also other works on the Penner model \ref\other{ C. Itzykson and J.
Zuber, Comm. Math. Phys. 134 (1990) 197\semi N. Chair and S. Panda, Phys.
Lett. 272B (1991) 230\semi N. Chair, Rev. Math. Phys. 3
(1991) 285\semi C.I. Tan, Mod. Phys. Lett. A6 (1991) 1373;
Phys. Rev. D45 (1992) 2862\semi M. Srednicki, Mod. Phys. Lett.A7 (1992)
2857\semi R. Brower, N. Deo, S. Jain and C.I. Tan, Harvard preprint,
HUTP-92/A035.}).

\newsec{KPZ Version of $c=1$}
In the last section we saw that the $SL(2,R)/U(1)$ Kazama-Suzuki model at
level $k=3$ is very similar to the $c=1$ system coupled to gravity, and it
becomes isomorphic to it if we set $\beta ={\sqrt \mu}$.  In this section
we will see that the correspondence is more transparent if we use the KPZ
formulation of non-critical matter coupled to gravity \kpz. In this
formulation, which can also be used to deal with the case where the
cosmological constant is not zero, the Liouville system is represented by
a {\it twisted} $SL(2,R)$ theory (whose energy momentum tensor has a term
proportional to $\partial J_3$) with $J^-$, having spin zero, set to a
constant equal to $\sqrt \mu$ \ref\shat{A. Alekseev and S. Shatashvili,
Nucl. Phys. B323 (1989) 719.}. This gauging introduces a pair of ghosts
$B,C$ of spins 1,0 with a KPZ reduction operator given by
$$Q_{KPZ}=B(J^- -{\sqrt \mu})$$
In addition to this, one adds the matter system and the diffeomorphism
$b,c$ ghosts. One then considers, on the reduced Hilbert space, the usual
string BRST operator
$$Q_{BRST}=c(T_m+T_l +{1\over 2} T_{ghost})$$
where the subscript $m$ refers to matter and $l$ refers to the combined
twisted $SL(2,R)$ and $B,C$ system. The level of $SL(2,R)$ that one needs
is fixed by the central charge of the matter \kpz , and in the case of
matter with $c=1$ it is given by $k=3$. Moreover, as pointed out in
\ref\marc{N. Marcus and Y. Oz, Preprint TAUP-1962-92, June 1992}, it turns
out that in the KPZ formulation we have to use a Wakimoto representation
for $SL(2,R)$ in order to get agreement with the usual $c=1$ theory. So
the full Hilbert space in the KPZ description for $c=1$ matter represented
by a scalar field $X$ is given by
$${\cal H}=[\phi]+[\beta ,\gamma ]+[B,C]+[X]+[b,c]$$
which is {\it exactly} the same Hilbert space we had in the KS case
discussed in the previous section, with the same expressions for the
energy momentum tensors (in both cases the $SL(2,R)$ is twisted and at
level $3$).

However, the BRST structure seems at first sight to be different: In the
KPZ case we first impose the constraint corresponding to the BRST charge
$$Q_{KPZ}=B(\beta -\sqrt{\mu} ) $$
This means that the $B,C,\beta, \gamma $ degrees of freedom can be ignored
(after setting $\beta = \sqrt{\mu}$), and we are left with $[\phi]+[X]
+[b,c]$ on which the physical subspace is obtained by considering the
string BRST $Q_{BRST}$. This gives the usual DDK-type expressions for the
physical states in terms of $X$ and $\phi$. In the KS theory we have
instead to reduce the Hilbert space first under
$$Q_{U(1)}=C(\beta \gamma - cb - \partial X^-)$$
and then impose the Kazama-Suzuki BRST
$$Q^+=c J^+=c(\beta \gamma^2 -{\sqrt 2} \gamma \partial \phi +3
\partial \gamma )$$
as we did in the previous section. As discussed there, the cohomologies
are represented by the same expressions as in $c=1$ theory, with some
extra $\beta$ dependence which disappeared when we set $\beta ={\sqrt
\mu}$. In other words, the cohomology states of the KPZ theory can be
chosen, by modifying with $Q_{KPZ}$ trivial fields, to have these extra
$\beta $ dependences and so can be chosen to coincide with those of KS
theory.  It is desirable to explain this coincidence at the level
of the BRST operators. The following shows this directly.

Note that in both the KPZ formulation and in the KS model
a twisted $N=2$ algebra is realized. The two supercharges of
the algebra in the KPZ case are $Q_{BRST}$ and $b_0$ and in the
KS case are $Q^+$ and $Q^-$. Moreover, we have
$$\{ Q_{BRST},b_0 \} = L_0$$
and
$$\{ Q^+, Q^-\} =L_0+\{ Q_{U(1)} ,\Lambda \}$$
where the last equality is just the statement of the twisted $N=2$
symmetry of the KS model. This symmetry is valid modulo the $U(1)$
constraint, explaining the existence of the second term for some $\Lambda
$. Now $Q^-$ differs from $b$ only because of an extra $\beta$, so using
the KPZ constraint we see that it is equivalent to $b_0$.

Consider now the subspace of the Hilbert space where the KPZ constraint
and the KS constraint are both satisfied. This is compatible in that if
we allow both positive and negative powers of $\beta$, any state in the
Hilbert space can be made to satisfy the KS constraint by dressing with
powers of $\beta$ without affecting the KPZ constraint, as $\beta$ is set
to a constant. On this reduced Hilbert space we see that the two BRST
charges $Q_{BRST}$ and $Q^+$, which are {\it defined} by the nilpotency
condition and the anti-commutation relations above, are equivalent, since
their defining relations are equivalent.

A more direct way of showing the equivalence of $Q_{BRST}$ and
$Q^+$, which was pointed out to us by V. Sadov (following a similar
computation done in \ref\sado{V. Sadov,{\it On the spectra of}
$sl(N)/sl(N)$-{\it cosets and} $W_N$ {\it gravities}, Harvard preprint
HUTP-92/A055 .}), is to note that
$$\beta^{-1}Q_{BRST}=Q^+ +\{ Q_{U(1)},\oint
cB(\gamma -
\beta^{-1}\partial X^+) +c \beta^{-1}\partial B\}$$
Thus the two BRST charges are equivalent upto a $Q_{U(1)}$-exact term.

To complete the proof that the (twisted) KS theory (coupled to topological
gravity) and the KPZ theory are really the same, we have to show that the
rules for computations of the correlation functions are also the same.  As
far as the screening operators are concerned, they are the same for both
KPZ and KS because they come from the same underlying $SL(2,R)$ theory.
The only other thing we need to check is the fact that in the KPZ case, as
in all string theories, we have to make $3g-3$ insertions of $b$ ghosts
(and similarly for the right-movers). In the context of twisted KS theory
coupled to topological gravity, this role is played by $3g-3$ insertions
of $G^-= b \beta $. This is the same as in KPZ when we use our freedom in
the KPZ theory to make $Q_{KPZ}$ trivial modification to $b$.  We have
thus shown that the twisted KS theory discussed in the previous section,
coupled to topological gravity, is equivalent to the $c=1$ theory coupled
to gravity in the KPZ formulation.

\newsec{Partition Function Computation}
In order to understand whether the above correspondence between the KS
theory and non-critical $c=1$ coupled to gravity is practically useful or
simply an elegant reformulation, we need to ask whether in the new
formulation the theory is amenable to explicit computation.  Remarkably
the answer to this question is in the affirmative and clearly demonstrates
the power of topological methods in this reformulation. In fact, this
particular KS theory coupled to topological gravity has been considered by
Witten \wittks\ who showed how to compute the partition function at
arbitrary genus, and certain correlation functions.

In \wittks\ it was found that if the level $k$ of the $SU(2)$ algebra is
continued to -3, the partition function of the twisted KS theory at
arbitrary genus (which is nonzero without any need for field insertions
because $\hat c=3$) is given by the Euler characteristic of the moduli
space of Riemann surfaces (This can be computed using a matrix model due
to Penner \pen.) On the other hand, it had already been discovered \DV\
that the partition function of the $c=1$ matrix model at the self-dual
radius, is given at arbitrary genus by the Euler characteristic of the
moduli space of Riemann surfaces. The reason for this correspondence
between the partition function of the KS model and that of the $c=1$
matrix model at self-dual radius was not understood, and was one of our
main motivations for undertaking this work. This is now explained by our
proof that the $c=1$ non-critical string is equivalent to the special KS
model. Thus we have demonstrated that the matrix model and continuum
results for the partition function of non-critical $c=1$ matter at the
self-dual radius are the same!

The Lagrangian formulation of the KS model, which is most suitable for
making use of topological field theory techniques in computing
correlations functions, has been discussed in \wittks, to which we refer
the reader for more details. Here we quote some of the main facts relevant
to our problem.  A subset of chiral fields, i.e., $Q^+$ invariant
fields, which exists for all $k$, was found to be concentrated on $SU(2)$
group configurations $g$ for which
\eqn\auone{g = \pmatrix{e^{i\theta} & 0\cr
0 & e^{-i\theta }\cr}   }
The chiral fields are represented by powers of $g_{11}$.  Let
$U_r=g_{11}^r $, and let us allow $r$ to be both positive and
negative\foot{The choice of negative powers of $g_{11}$ is more natural
if we consider the group to be $SU(1,1)$ (since the level is negative in
$SU(2)$ language) as in this group, $g_{11}$ is never zero.}.
We will discuss below how these fields are related to the tachyon fields
discussed in section 3. The result for the $n$-point function of these
operators at arbitrary genus is given by \wittks
\eqn\coran{\langle U_{r_1}U_{r_2}...U_{r_s} \rangle_g =\int_{{\bar {\cal
M}}_{g,s}}
c_T({\cal \nu )}}
where ${\bar {\cal M}}_{g,s}$ denotes the compactified moduli
space of Riemann surfaces of genus $g$ with $s$ punctures, $\cal \nu$ is
a bundle over this space whose fibers are given by
$$V=H^0(\Sigma , K^{\otimes 2}\otimes_i O(z_i)^{-r_i})$$
where $\Sigma$ is a Riemann surface of genus $g$ and $K$ is its canonical
line bundle, and $c_T$ in \coran\ denotes the top chern class of the
bundle. With no $U$'s inserted, \coran\ gives the Euler characteristic of
${\bar {\cal M}}_g$.

To understand the significance of the $U_r$ fields, note that $g_{11}$ is
a HWS (because it is the $11$ component of the matrix field).  In fact,
$g_{11}^r$ will belong to the highest weight state of the representation
$D^-_{r/2}$, which should thus correspond to the tachyon states $V_{r/2}$
defined in Eq.\tachhws. In particular, as discussed there, $V_{-1/2}$,
(which comes from $r=-1$) corresponds to the cosmological constant
operator. Using \coran\ we can compute the s-point function of the
cosmological constant operator because in that case the Euler character of
$\cal \nu$ with $V=H^0(\Sigma , K^{\otimes 2}\otimes_i O(z_i))$ is the
same as the Euler character of ${\bar {\cal M}}_{g,s}$. From \DV\ this is
known to be the same answer as that of the matrix model with
$s$ insertions of the cosmological constant operator.  So we find further
confirmation that the continuum and matrix model results match.

There is a conservation law in \coran\ which implies \wittks
\eqn\concor{\sum_i (r_i+1)=0}
This conservation law is independent of the genus $g$. In fact
it is exactly what we would expect given the identification of
$U_r$ with a HWS with $j=r/2$, because we have from the identification
of section 3, that the $X$--momentum $k$ (up to a factor of $1/\sqrt{2}$) is
given by \tachhws:
\eqn\matmo{k=2j+1=r+1}
Therefore \concor\ is equivalent to
$$\sum_i k_i=0$$
This is just matter momentum conservation, which holds in
every genus.

Let us now turn to the computation of the 2,3 and 4 point functions
of the tachyon operators $U_r$ on the sphere.  These computations
have also been performed in \wittks (all we need to do is make
a simple modification in the case of 4 point functions on the sphere):
To compute the two point function, it is more natural to first compute the
three point function, and then set one of the operators to be
the cosmological constant operator.
The three-point function on the sphere is (eq. 3.36 of \wittks )
$$\langle U_{r_1} U_{r_2} U_{r_3} \rangle_{g=0}=\delta_{(r_1+r_2+r_3+3)}$$
which using \matmo\ is simply the statement that the three point function
is 1 if the matter momentum conservation is satisfied and zero otherwise.
To obtain the two point function, we simply need to set one of the $r_i$
to -1, which again implies that the two point function is 1 if the two
momenta are equal and opposite. (We are taking the cosmological constant
to be equal to 1, but the right powers of $\mu$ can always be restored by
considering the amount of momentum violation for the $\phi$ field).

Now we turn to 4-point tachyon correlators on the sphere.
The four point computation in the KS model has been done for positive
$SU(2)$ level in \wittks. This can be adapted to our case by a slight
modification in Eq.(4.62) of that paper (with $\gamma =-1$):
\eqn\forpo{\langle U_{r_1}U_{r_2}U_{r_3}U_{r_4}\rangle =
[{-1\over 2}{\rm max}(r_1+r_2 ,r_3+r_4) -
{2\over 3}]+(r_2 \leftrightarrow r_3)
+(r_2\leftrightarrow r_4).}
Using $k=r+1$ and $\sum k_i =0$ this is equal to\foot{This result is also
valid when one considers rational radii by going to the $n$-fold cover of
$SU(1,1)$, and agrees with the standard results of $c=1$ theory.}
\eqn\fifor{\langle U_{r_1}U_{r_2}U_{r_3}U_{r_4}\rangle =
-{1\over 2}|k_1+k_2|-{1\over 2}|k_1+k_3|-{1\over 2}|k_1+k_4|+1}
which is equal to the four point tachyon amplitude on genus zero computed
using either the matrix model \ref\moore{G. Moore, R. Plesser and S.
Ramgoolam, Nucl. Phys. B377 (1992) 143.}
or the DDK approach \dikut\
(with the leg factors absorbed into the definition of $U_r$).

We have thus found that the partition function to all genus,
the $n$-point function of the cosmological constant to all genus,
and the $n$-point function of tachyons up to $n=4$ at genus zero,
can all be explicitly computed in the KS model, and agree with the
corresponding matrix model results.

Since we know explicitly the $n$-point tachyon correlators in the matrix
model \moore \dijmoor\ and there is an expression (given in Eq.\coran) for
the corresponding amplitudes in the KS theory, we have a conjectured
equality between these quantities to all genus and for all $n$.

We have been somewhat imprecise as to which chirality tachyon amplitude we
are considering as there is a puzzle here.  From the naive relation we
discussed above for the relation of $U_r$ with the HWS tachyon state with
$j=r/2$, the two and three point function computation seem to be in
contradiction with the computation using explicit representations we
obtained in section 3.  In particular two HWS tachyon fields do not have a
non-vanishing two point function.  We need the dual field which
corresponds to a tachyon of opposite chirality which includes a factor
$\beta^{-1}$ in order to give a non-zero answer.  The same is true for the
three point functions.

We believe that the clue to resolving this puzzle is to note the role of
the screening current \scre. Indeed this suggests that $g_{11}^r$ for
$r>0$ is to be identified with the dual tachyon fields which have extra
factors of $\beta^{-1}$ and are of opposite chirality.  The way the two
point function comes out to be non-vanishing in this case is by using the
screening current \scre\ to take care of $\phi$ momentum conservation.
Indeed, application of an appropriate number of \scre\ currents to
tachyons of one chirality seems to map them to tachyons of the other
chirality, by replacing a negative Liouville dressing with a positive one.
This is expected in view of the fact that we are at a non-zero
cosmological constant and the Liouville momentum conservation is not
valid, or more precisely it is valid only when we also count the factors
of the cosmological constant. Therefore, the fact that only one set of
tachyons appears in \coran\ should be related to the fact that for each
matter momentum there is a unique state with negative Liouville dressing
which should be identified with $U_r$.

As we know, the $c=1$ model has, in addition to tachyon states, an
infinite number of discrete states.  As we discussed in section 3, they
also appear in the KS theory.  How do we compute correlation functions
involving some discrete states using topological techniques?  This is a
very important question to which an answer needs to be found.  There is
also a complementary question:  We know that for each tachyon field $U_r$
there is an infinite series of gravitational descendants $\tau_n (U_r)$
for which the abstract definition of correlation function, a
generalization of \coran\ is given in \wittks:
$$\langle \tau_{n_1}(U_{r_1})\tau_{n_2}(U_{r_2})...\tau_{n_s}(U_{r_s})
\rangle =\int_{{\bar {\cal M}}_{g,s}}c_T({\cal \nu} )\cdot \prod_{i=1}^s
(c_1({\cal L}_{(i)}))^{n_i}$$
where $c_1({\cal L}_{(i)})$ denotes the first chern class of the
line bundle on moduli space whose fiber is the space of one-forms at the
puncture on the Riemann surface.  The selection rule now gets modified to
$$-\sum_i(r_i+1) +\sum_i n_i=0$$
One is tempted to identify the extra states of the topological theory
(the gravitational descendants) with the missing states of KS theory
(the discrete states).  It seems that the most natural description of the
topological descendants may involve using the dual Wakimoto module discussed
in section 3, where the topological gravity multiplet
is automatically available.
This does not appear to be completely
straightforward, however, and this problem is under investigation.

One may also ask if one can use topological techniques to compute
correlation function of the $c=1$ theory away from the self-dual point. In
principle this must be possible because following the logic of section 3,
we see that going to the $n$--fold cover of $SU(1,1)$, the $X$--field will
live on a circle whose radius is $n$ times the self-dual radius.  It
should be possible to translate this geometric statement into a
computation of the partition function\foot{One might formally think that
this should give as partition function the Euler characteristic of the
moduli space of Riemann surfaces with a $Z_n$ flat connection, but this
only modifies the answer from the self-dual result by the factor $n^g$ at
genus $g$, whereas the expected answer from the $c=1$ matrix model is a
polynomial in $n$ and $n^{-1}$, invariant under $n\leftrightarrow n^{-1}$
with the highest power $n^g$.} by carefully finding the modifications
needed in \wittks\ when one goes to the multiple cover.

\newsec{Black Hole Interpretation}
We have seen in the previous sections that $c=1$ non-critical string is
equivalent to a superconformal $SL(2,R)/U(1)$ coset.  Let us compare this
with a proposal in \bwit\ where it was suggested that non-critical $c=1$
string may be related to an $SL(2,R)/U(1)$ background.

In fact, the construction of section 3 is very reminiscent of similar
considerations in the context of the black hole of
\bwit\ done in \ref\beretal{R. Dijkgraaf, E. Verlinde and H. Verlinde,
Nucl. Phys. B 371 (1992) 269\semi
M. Bershadsky and D. Kutasov,
Phys. Lett. B266 (1991) 345\semi
J. Distler and P. Nelson, Nucl. Phys. B 374 (1992) 123\semi
T. Eguchi, H. Kanno and S.-K. Yang,
$W_\infty$ {\it Algebra in two-dimensional black hole,}
Preprint NI92004\semi H. Ishikawa and M. Kato, {\it Equivalence of BRST
cohomologies for 2-d black hole and c=1 Liouville theory}, preprint,
UT-Komaba/92-11.}. In that case, with the choice of level $k=9/4$, the
central charge of the $SL(2,R)$ is 27. Taking a $U(1)$ coset decreases
the central charge by 1 to give $26$ and then the ghost system $b,c$ with
central charge $-26$ is added, to get total central charge zero: %
\eqn\bla{{SL(2,R)\big|_{27}\over U(1)}\otimes [b,c]}
Our supersymmetric version, on the other hand, can be described in the
following way: we start with an $SL(2,R)$ at level $k=3$ with twisted
central charge $27$ (twisted in the sense that we have added a term
proportional to $\partial J_3$ to the energy momentum tensor), add to it a
$b,c$ system with central charge $-26$, giving a net central charge of 1,
and then take a $U(1)$ quotient which decreases the central charge by 1,
to give  zero net central charge:
\eqn\subl{{SL(2,R)\big|_{27} \otimes [b,c]\over U(1)}}

There are two differences between these two theories. One is that the
supersymmetric model that we study contains a twisted $SL(2,R)$ theory,
which is why the two theories have the same central charge $27$ at two
different levels $k=9/4$ and $k=3$. The second is that in taking the
$U(1)$ coset in the case of \bla\ there is no mixing with the $[b,c]$
system, whereas for us, the $U(1)$ in \subl\ mixes the `matter' and the
`ghost' system, as its generator is given by $J_3+bc$.

The question is, which background does our theory (given by \subl)
describe? In fact this has already been studied in Refs.\ref\bar{I. Bars,
{\it String propagation on black holes}, preprint USC-91-HEP-B3\semi
Nucl. Phys. B 334 (1990) 125.}\ref\Eguc{T. Eguchi, Mod. Phys. Lett. A7 (1992)
85.}\ where it is concluded that the supersymmetric coset
also describes the same black hole configuration!  So we are led to a
puzzle:  If we are interested in describing bosonic strings propagating in
a black hole background, which coset is the correct choice:
\bla\ or \subl ?

Based on experience with static string solutions, one might be inclined to
pick \bla\ as the correct choice, as the ghost and matter systems are
always treated independently, whereas \subl\ mixes the two. However, one
expects that for time dependent backgrounds {\it there should be a mixing
between the matter and the ghosts}. There are various ways to see this.
Perhaps the simplest indication that this might happen is in the proof of
the no-ghost theorem for bosonic strings given in \ref\fgz{I. Frenkel, H.
Garland and G. Zuckerman, Proc. Nat. Acad. Sci. USA 83 (1986) 8442.}. It
is crucial in their proof that the signature character of the ghost
Hilbert space cancels part of the matter signature character given by an
uncompactified flat 1+1 dimensional spacetime. So if we change the matter
theory such that there is no 1+1 dimensional uncompactified flat spacetime
left, then there is bound to be some change in our ghost system as well,
to maintain the no-ghost theorem. Therefore for the 1+1 dimensional black
hole geometry one expects a mixing, and indeed \subl\ is a very ``mild''
mixing of the matter system with the ghost system, via a $U(1)$
constraint. The fact that the matrix model partition function agrees to
all genus with the partition function of \subl\ is, in our opinion, a
strong indication that the correct theory describing bosonic string
propagation in the black hole background is in fact \subl\ and not
\bla.

This also suggests that maybe {\it the time dependent backgrounds are
described in bosonic string theory by a twisted} $N=2$ SCFT {\it as is the
case here}. This is a very interesting question for further research,
even in the context of critical strings.

Given this connection between matrix models and black holes, it is natural
to ask what we may learn about black hole singularity. So far we have been
talking about the Euclidean black hole. The Minkowski version is obtained
by changing the $U(1)$ imbedding \bwit. In our case, which has an $N=2$
supersymmetry, there is an observation in \Eguc\ that the simplest chiral
fields (those given by $g_{11}^r$) correspond to the points on the black
hole singularity! This is very significant as it implies that the
singularity itself encodes all the physical degrees of the freedom of the
black hole and that it is a topological degree of freedom, a notion which
was previously suggested in \bwit. Perturbing the theory by throwing in
tachyons is equivalent to perturbing the singularity structure, namely,
changing the topological $N=2$ theory by topologically relevant
perturbations.

Another question to ask is whether string theory makes the singularity go
away, or more precisely what is the fate of the singularity. This is a
more difficult question to answer.  In order to discuss this point let us
first consider the similar but simpler situation of orbifold singularities
\ref\orb{L. Dixon, J. Harvey, C. Vafa and E. Witten, Nucl. Phys. B274
(1986) 285.}. In that case also, the singularities can be viewed as
topological degrees of freedom of the $N=2$ theory, as was discussed in
\ref\morb{S. Cecotti and C. Vafa, Mod. Phys. Lett. A7 (1992) 1715.}(see
also \ref\zas{E. Zaslow, {\it Topological orbifold models and quantum
cohomology rings}, preprint, HUTP-92/A065.}).  In fact the orbifold
singularities contribute to Witten's index $Tr(-1)^F$. In the case of
orbifolds it is well known that singularities are harmless for string
propagation \ref\intr{S. Hamidi and C. Vafa, Nucl. Phys. B279 (1987)
465\semi L. Dixon, D. Friedan, E. Martinec and S. Shenker, Nucl. Phys.
B282 (1987) 13.}\ and one does not need to get rid of them to make the
theory consistent, even though if we were describing point particle
propagation in the presence of orbifolds we would get an inconsistent
theory.  For strings not only it is not needed to get rid of the
singularity, but in fact any {\it finite} perturbation of the theory will
leave a trace of the singularity, because $Tr(-1)^F$, being an index,
cannot change.  Perturbation will resolve the singularity but will still
lead to the same contribution to Witten index.

The above situation is very similar to that of black holes.  In the case
of black holes also, the points of singularity contribute to $Tr(-1)^F$,
as they are topological.  Moreover being described by a conformal coset
theory suggests that the strings have a non-singular propagation on the
black hole\foot{This needs to be verified by considering correlations of
the Minkowskian coset theory.}. To see the fate of the black hole, we have
to consider string loop corrections in the Minkowskian version. As we
discussed, the matrix model is equivalent to the Euclidean black hole and
it is thus not clear how the computations in the Euclidean set up will
have to be interpreted for the Minkowskian version.  But since the black
hole singularity is topological, whatever happens to the singularity it
cannot quite disappear, because it contributes to a supersymmetric index!

The Minkowskian version of the Kazama-Suzuki model will have to be studied
in more detail to shed light on aspects of the black hole. It is amusing
that even though the matrix model is equivalent to Euclidean black hole,
it involves `scattering' for tachyons. This is because one allows
exponentially growing (or decaying) solutions in one of the Euclidean
directions. It would be
interesting to connect the scattering matrix thus obtained to that for the
Minkowskian black hole.

At any rate, it is quite remarkable that with the dictionary we have
found, the partition function of the Euclidean 2d black hole can be
computed for all genus, and is given by the Euler characteristic of the
moduli space of Riemann surfaces of that genus.

\newsec{Generalizations and Conclusions}
We have demonstrated that $c=1$ matter coupled to gravity with non-zero
cosmological constant is equivalent to a topological superconformal
$SL(2,R)/U(1)$ coset coupled to topological gravity. This solves the
problem of finding a topological field theory for the $c=1$ string, and
leads to a systematic computation of the continuum theory correlations
using topological techniques. All the results that have been checked
(including correlation of arbitrary number of cosmological constant
operators at all genus and up to four-point tachyon scattering amplitude
on genus zero) agree with those of the $c=1$ matrix model. Moreover, we
have shown how this theory is related to a bosonic string propagating in a
two dimensional black hole background.

There are a number of directions that are suggested by these observations:
We now have an exact, completely solvable representation of the Euclidean
black hole in terms of the $c=1$ matrix model. What does this teach us
about black holes? How can we use similar ideas to describe
other time-dependent string solutions?
 We have discussed some aspects of these questions in
section 6.

Another question worth investigating is to show explicitly that the
results of the matrix models tachyon scattering amplitudes for all genus
given in \moore\ agree with that predicted by this correspondence and
given by \coran. Turning things around, we thus have a Kontsevich-like
representation of the invariants \coran\ given by a Penner-like matrix
integral as obtained in \dijmoor .

Finally, it is natural to ask if the above ideas can be generalized to
other models with $c<1$. The most obvious generalization would be just to
change the level of the KS coset model. Even though this has been used in
conjunction with topological gravity beginning with the work of Li
\ref\lire{K. Li, Nucl. Phys. B354 (1991) 711.}, it is not in the same
spirit as what we are doing here in that the ghost system is {\it part} of
the matter system for us and we do not have to add it by hand.  In
particular as is clear from \spins, only for $k=3$ do we get the $b,c$
ghost system with the correct spins.

Instead it turns out that a remark by Frenkel (see appendix A), that the
superconformal KS cohomology computation at $k=3$ is equivalent to the
{\it bosonic} $SL(2,R)/SL(2,R)$ at the same level, can be used to
generalize the above construction \bsadv. Indeed it was already known
that the physical states of $SL(2,R)/SL(2,R)$ topological theory are
isomorphic to those of matter coupled to gravity \ref\gmodg{M. Bershadsky
and E. Frenkel, unpublished\semi O. Aharony, O. Ganor, N. Sochen, J.
Sonnenschein and S. Yankielowicz, {\it Physical states in} $G/G$ {\it
models and 2d gravity}, Preprint TAUP-1961-92\semi O. Aharony, J.
Sonnenschein and S. Yankielowicz, $G/G$ {\it models and} $W_N$ {\it
Strings}, Preprint TAUP-1977-92\semi
H.L. Hu and M.Yu, {\it On BRST cohomology of}
$SL(2,R)/SL(2,R)$ {\it gauged WZNW models}, Preprint AS-ITP-92-32\semi P.
Bouwkneght,J. McCarthy and K. Pilch, {\it Semi-infinite cohomology in CFT
and 2d gravity}, Preprint CERN-TH. 6646/92.}\sado. However there are two
new ingredients here which make this work: first, one should take a {\it
twisted} bosonic $SL(2,R)/SL(2,R)$ theory, and secondly it turns out that
by a careful reorganization of the BRST, one obtains an $N=2$
superconformal algebra for this theory \bsadv. The twisting has the
effect of making one pair of the gauge ghosts act as the $b,c$ system with
the correct spin, and the other two pairs act as KPZ reduction \kpz\ and
Hamiltonian reduction \ref\beroo{M. Bershadsky and H. Ooguri, Comm. Math.
Phys. 126 (1989) 49.}\ of the rest of the degrees of freedom (to Liouville
and matter respectively). The possible connections of this twisted
$SL(2,R)/SL(2,R)$ theory to three dimensional gravity are also currently
under investigation \bsadv .

\vglue 2cm
We would like to thank M. Bershadsky, D. Ghoshal, C. Imbimbo, D. Jatkar,
S. Mathur, G. Moore, H. Ooguri, V. Sadov, E. Verlinde, H. Verlinde and E.
Witten for valuable discussions. We are greatly indebted to E. Frenkel for
taking up the question of computation of the cohomology of the KS coset
model presented in appendix A.

S.M. would like to acknowledge the kind hospitality of the relevant
departments at Harvard University, Princeton University, Ecole Normale
Superieure, and INFN, Genova, where part of this work was done.

The research of C.V. was supported in part by Packard Foundation and
NSF grants PHY-89-57162 and PHY-87-14654.

\def\lo{\longrightarrow}
\def\blo{\longleftarrow}
\def\ri{\rightarrow}
\def\le{\leftarrow}
\def\bul{\bullet}

\appendix{A}{Cohomology of Kazama-Suzuki Model}
\medskip
\centerline{{\bf by Edward Frenkel}}
\medskip

In this Appendix we will compute the spins of the cohomology states of the
Kazama-Suzuki (KS) model (cf. Section 3).

For a module $M$ over the affine Kac-Moody algebra $\hat{SL}(2)$ of level
$-3$ let us consider the complex $M \otimes [X] \otimes [b,c]
\otimes [C,B]$, graded by the ghost number, with respect to the action of
the differential $$\int [C(J^3 - i{\partial X\over {\sqrt 2}}) + cJ^+ -
Ccb] dz.$$ Let us denote by $H^q(M)$ the space of cohomology states of this
complex (relative to the 0th mode of $C$), of ghost number $q$. If $M$ is
the Wakimoto module $W_j$ with HWS $D^-_j$,
with $j$ integral or half-integral, then $H^q(W_j)$ is exactly the
space of states of ghost number $q$ of the KS model, of ghost number $q$.
We will write $H^q(M) =
\{j_1,\ldots,j_n\}$, if $H^q(M)$ is linearly spanned by states of spins
$j_1,\ldots,j_n$, and will use the notation $S_{j_1,j_2}$ for the set
$\{j_1,j_1-1,\ldots,j_2+1,j_2\}$.

\noindent {\bf Theorem} {\it
(a) If $j=-1/2$, or $j=-1$, then $H^1(W_j) = \{ j\}$;

(b) If $j\geq 0$, then $H^1(W_j) = S_{j,-j-1}, H^2(W_j) = S_{j-1,-j-1}$;

(c) If $j<-1$, then $H^1(W_j) = S_{-j-2,j}, H^0(W_j) = S_{-j-2,j+1}$;

and all other $H^q(W_j) = 0$.}

To prove this Theorem, we will use the following result on the structure of
the Wakimoto modules with negative levels, which was proved in \frenko\
(where we used the opposite notations for Wakimoto and dual Wakimoto
modules): for $j \geq
-1/2$ the module $W_j$ is isomorphic to the Verma module with the same HWS,
and for $j<-1/2$ the module $W^*_j$ is isomorphic to the Verma module with
the same HWS.

This result is quite surprising. Indeed, the Verma modules are defined as
the modules, freely generated from their HWS under the action of the
subalgebra of $\hat{SL}(2)$, spanned by $J^3_n, n<0, J^+_n, n<0,$ and
$J^-_n, n\neq 0$, whereas the Wakimoto modules $W_j$ (or dual Wakimoto
modules $W^*_j$) are defined by explicit embeddings of $\hat{SL}(2)$ into
the free fields. These embeddings are such that the modules $W_j$ ($W^*_j$)
are free over the Lie subalgebra, spanned by $J^3_n, n<0$ and $J^-_n, n\leq
0$ ($J^3_n, n<0$ and $J^+_n, n<0$), and co-free over the Lie subalgebra,
spanned by $J^-_n, n>0$ ($J^+_n, n\geq 0$) (cf. \ref\ff{B.Feigin, E.Frenkel,
Comm. Math.  Phys. 128 (1990) 161}).

Now, the structure of Verma modules over $\hat{SL}(2)$ is known from
\ref\ffk{B.Feigin, E.Frenkel, in Physics and Mathematics of Strings,
V.Knizhnik Memorial Volume, eds. L.Brink, e.a., 271-316, World Scientific,
1990}, and this allows us to describe the structure of $W_j$ and $W^*_j$.

Our computation is based on the following result.

\noindent {\bf Lemma} {\it
For the dual Wakimoto module $W^*_j$ we have: $H^1(W^*_j) = \{j\}$, and
$H^q(W^*_j)=0$ for $q \neq 1$.}

This follows from the properties of $W^*_j$ with respect to the action of
the operators $J^3_n$ and $J^+_n$, described above.

The modules $W_{-1}$ and $W_{-1/2}$ are irreducible and therefore they are
isomorphic to $W^*_j$. Hence, part (a) of the Theorem follows from the
Lemma. We can prove the rest of the Theorem by induction, separately for
integral and half-integral values of $j$.

We will demonstrate it here in the case of integral spins. The case of
half-integral spins can be pursued along the same lines.

The modules $W_j$ with $j\geq 0$ are isomorphic to Verma modules and have
the structure:
\eqn\str{\bul^{I_j} \lo \bul^{I_{-j-1}} \lo \bul^{I_{j-1}} \lo \ldots
\bul^{I_0} \lo \bul^{I_{-1}}.}

The modules $W_j$ with $j<0$ are isomorphic to dual Verma modules and
have the structure:
\eqn\strr{\bul^{I_j} \blo \bul^{I_{-j-2}} \blo \bul^{I_{j+1}} \blo \ldots
\bul^{I_0} \blo \bul^{I_{-1}}.}

Here $I_j$ denotes the irreducible module with HWS $D^-_j$. These diagrams
have the following meaning: the dots represent the irreducible subfactors
of $W_j$, and the arrows show, which way we can get from one of them to
another.

The diagram for the module $W^*_j$ can be obtained from the diagram for the
module $W_j$ by reversing the arrows.

Assume that we have already proved the Theorem for all integral values of
$j$, such that $|j+1| \leq n$, where $n\geq 0$ is an integer.

In particular, the diagram \strr\ with $j=-n-1$ and the reverse to the
diagram \str\ with $j=n$ imply that $I_n$ is a submodule of $W^*_n$,
and the quotient is isomorphic to $W_{-n-1}$. Therefore, the following exact
sequence holds:
\eqn\es{0 \ri I_n \ri W^*_n \ri W_{-n-1} \ri 0.}

Recall that every exact sequence of modules $0 \ri A \ri B \ri C \ri 0$
leads to the following exact sequence of cohomologies
$$\ldots \ri H^q(A) \ri H^q(B) \ri H^q(C) \ri H^{q+1}(A) \ri \ldots.$$

Our sequence \es\ gives: \eqn\ess{0 \ri H^0(W_{-n-1}) \ri H^1(I_n) \ri
H^1(W^*_n) \ri H^1(W_{-n-1}) \ri H^2(I_n) \ri 0,} because all other
cohomologies are 0. By our induction assumption and the Lemma, we have:
$H^1(W^*_n) = \{n\}, H^1(W_{-n-1}) = S_{n-1,-n-1}$. Therefore, there can be
no non-trivial map between them, and the sequence \ess\ splits into two:
\eqn\per{0 \ri H^0(W_{-n-1}) \ri H^1(I_n) \ri H^1(W^*_n) \ri 0} and
\eqn\vtor{0 \ri H^1(W_{-n-1}) \ri H^2(I_n) \ri 0.}

According to \per, $H^1(I_n) = H^0(W_{-n-1}) + H^1(W^*_n) = S_{n,-n},$
and according to \vtor, $H^2(I_n) = H^1(W_{-n-1}) = S_{n-1,-n-1}$, and also
$H^q(I_n) = 0$ for $q \neq 1,2$.

Further, we have the exact sequence $$0 \ri W^*_{-n-1} \ri W_n \ri I_n \ri
0.$$ By the same method, we obtain $H^1(W_n) = H^1(W^*_{-n-1}) + H^1(I_n) =
S_{n,-n-1}, H^2(W_n) = H^2(I_n) = S_{n-1,-n-1}$, and $H^q(W_n) = 0$ for $q
\neq 1,2$.

Now, using this result and the exact sequence $$0 \ri W_n \ri W^*_{-n-2}
\ri I_{-n-2} \ri 0,$$ we obtain in the same fashion: $H^1(I_{-n-2}) =
S_{n-1,-n-2}, H^0(I_{-n-2}) = S_{n,-n-1}$, and $H^q(I_{-n-2}) = 0$ for $q
\neq 0,1$. This, together with the exact sequence $$0 \ri I_n \ri W_{-n-2}
\ri W^*_n \ri 0,$$ gives: $H^1(W_{-n-2}) = S_{n,-n-2}, H^0(W_{-n-2}) =
S_{n,-n-1}$, and $H^q(W_{-n-2}) = 0 $ for $q \neq 0,1$.

Thus, we have now proved our statement for integral values of $j$, such
that $|j+1| \leq n+1$. This completes the proof of the Theorem.

\noindent {\it Remarks}

(1) In the course of proving our Theorem, we have computed the cohomology
$H^q(I_j)$, corresponding to the irreducible representations $I_j$. This
cohomology is the same as that of $W_j$, except that one state is missing
in $H^1(I_j)$: of spin $-j-1$ for $j \geq 0$, and of spin $-j-2$ for $j
\leq -1$.

(2) It is known that $W_j$ is irreducible, if $j$ not integral or
half-integral. In that case $W_j$ is isomorphic to $W^*_j$, and therefore
the only cohomology class, which occurs, is the HWS.

(3) The technique that we have described will work for other negative
rational levels as well. We can again use the fact that $W_j$ are
isomorphic to Verma modules for $j\geq -1/2$, and $W^*_j$ are isomorphic to
Verma modules for $j<-1/2$ \frenko. However, in general, the structure of
Verma modules can be more complicated: we may have not one, but two strings
of singular vectors (cf. \ffk,\frenko).

(4) We would like to remark that at negative rational levels there are no
good projections from Wakimoto modules to irreducible modules, in contrast
to irrational, or positive rational levels.

For irrational levels, the irreducible representation of integral or
half-integral spin $j\geq 0$ can be defined as the 0th homology of the
complex \eqn\res{0 \le W_j \le W_{-j-1} \le 0,} or as the 0th cohomology of
the complex \eqn\resd{0 \ri W^*_j \ri W^*_{-j-1} \ri 0.} In these complexes
the only non-trivial differential is a multiple integral of the $(2j+1)$st
power of the corresponding screening current (cf. formula \scre).

For positive rational levels, we should extend the complexes \res\ and
\resd\ to a two-sided resolution as explained in \ff,\ref\bfel{D.Bernard,
G.Felder, Comm. Math. Phys. 127 (1990) 145}. The resulting complexes
will have the property that all but the middle cohomology vanishes, and the
middle cohomology is isomorphic to the irreducible representation.

At rational negative levels (for example, $k=-3$) the complexes
\res\ and \resd\ no longer work, because the power of the screening charge
is equal to 0 in this case. On the other hand, the complex
\es\ that we used in the proof of our Theorem, includes both Wakimoto and dual
Wakimoto modules, and therefore the corresponding differential can not be
expressed in a local form.

We may also consider the following complex \eqn\nov{0 \le W_j \le W_{j-1}
\le 0,} (for $j\geq 0$) with the screening charge \scre\ as the
differential. However, the analysis of the structure of the Wakimoto
modules, given by the diagram \str, shows that the 0th homology of this
complex is not the irreducible module, but a module $X_j$, which has the
following structure: $$\bul^{I_j} \lo \bul^{I_{-j-1}}.$$ It contains two
irreducible factors: $I_j$ and $I_{-j-1}$. Of course, we can introduce a
resolution of $I_j$, consisting of the modules $X_n$, but it can not be
made into a resolution, which consists of the Wakimoto modules.

(5) Finally, let us remark that the cohomology states of KS model at $k=-3$
are in one to one correspondence with the cohomology states of bosonic
$SL(2)/SL(2)$ theory at level $-1$. Indeed, let us consider the complex of
(relative) BRST cohomology of the whole algebra $SL(2)$ with coefficients
in $W_j \otimes W_{j'}$, where $W_{j'}$ is the Wakimoto module of level
$-1$ with HWS of spin $j'$ (cf. \gmodg). Then, by the properties of the
Wakimoto modules with respect to the action of the operators $J^-_n$,
described above, the cohomology of this complex reduces to spin $-j'-1$ sector
of the cohomology of KS model. Indeed, we can strip off the bosonic ghosts,
involved in the module $W_{j'}$, together with the fermionic ghosts,
corresponding to the current $J^-$, without changing the cohomology.

The cohomology classes of KS model, constructed in Section 3, can then be
viewed as symbols of the cohomology classes of the $SL(2)/SL(2)$ model with
respect to a certain filtration.

\listrefs
\end